%% file: main.tex
\documentclass[titlepage,rmp,twocolumn,hidelinks,10pt,byrevtex,secnumarabic,nofootinbib,longbibliography,a4paper,showkeys,preprintnumbers]{revtex4-2}
\usepackage{graphicx}
\usepackage{ccicons}
\usepackage{amsmath}
\usepackage{amssymb}
\usepackage{float}
\usepackage{censor}
\usepackage{moreverb}
\usepackage{url}

\usepackage{xr-hyper}
\usepackage{xcolor}
\usepackage{colortbl}
\usepackage[draft,commandnameprefix=always]{changes}

\usepackage{tikz}
\usepackage{pgfplots}
\usepackage{listings}

\input{zdoi.tex}
\input{variables.tex}

\newcommand{\zenodo}[1]{\doi{10.5281/zenodo.#1}}
\usetikzlibrary{shapes,arrows,arrows.meta,positioning,calc,external,patterns}
\usepgfplotslibrary{fillbetween}
\pgfplotsset{compat=1.11}

\usepackage[pdftex,
pdfauthor={Jose-Maria Martin-Olalla},
pdftitle={Does classical thermodynamics need a third law? Securing the second law at absolute zero},
            pdfsubject={Foundations of Thermodynamics},
            pdfkeywords={second law of thermodynamics; third law of thermodynamics; specific heat; statistical mechanics; entropy; temperature; Carnot engine; foundations of thermodynamics; Carnot theorem; absolute zero},
            pdfproducer={Latex with hyperref, or other system},
            pdfcreator={pdflatex, or other tool}]{hyperref}

                        \newcommand\rorlink[2]{\href{https://ror.org/#2}{#1\,\includegraphics[scale=0.1]{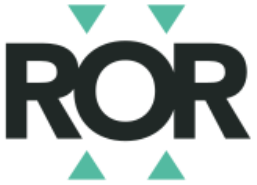}}
}
\usepackage{orcidlink}
\providecommand{\orcidlinki}[2]{\href{https://orcid.org/#2}{#1}\orcidlink{#2}}
            \usepackage{breakurl}
\usepackage{siunitx}
\usepackage{mathtools}
\usepackage[utf8]{inputenc}
\usepackage[english]{babel}
\usepackage{lineno}

\AtBeginDocument{\usepackage{booktabs}}

\graphicspath{{../}}

\makeatletter\AtBeginDocument{\let\@elt\relax}\makeatother

\newcommand{\uaddress}{\rorlink{Universidad de Sevilla}{03yxnpp24}, Facultad de Física, Departamento de Física de la Materia Condensada,  ES41012 Sevilla, Spain}

\begin{document}
\author{\orcidlinki{José-María Martín-Olalla}{0000-0002-3750-9113}}
\affiliation{\uaddress}
\email{olalla@us.es}
\thanks{\ccby}

\homepage{https://twitter.com/MartinOlalla\_JM}

\title{Does classical thermodynamics need a third law? Securing the second law at absolute zero}

\published[This version available ]{June 28, 2026}

\preprint{\textcolor{blue}{This is a preprint posted in \texttt{Zenodo} under \zenodo{\zdoi}}}

\begin{abstract}
  \input{abstract.tex}

    This version is published in \texttt{Zenodo} \zenodo{\zdoi}.

    Total word count: \texttt{\input{mainCount.tex}}

  \tableofcontents
\end{abstract}

\input{keywords.tex}

\maketitle

\input{manuscript.tex}

\input{main.bbl}

\input{history.tex}

\end{document}

%% file: zdoi.tex
\newcommand{\zdoi}{20745178}

%% file: variables.tex
\newcommand{\ucadena}[2]{%
        \ifisesp
        \else
        \fi}
\newcommand{\uQ}{\ensuremath{Q}}
\newcommand{\uW}{\ensuremath{W}}

\newcommand{\Qrev}{\ensuremath{\mathsf{\uQ}}}
\newcommand{\Wrev}{\ensuremath{\mathsf{\uW}}}

\newcommand{\ut}{\ensuremath{T}}

\providecommand{\equilibrio}{\ensuremath{\mathrm{E}}}

%% file: abstract.tex
This paper elaborates on the formal implications of the relationship between the Second and Third Laws and provides a comprehensive formal and historical justification for the logical redundancy of the Nernst heat theorem. By revisiting the Nernst-Einstein debate through the axiomatic frameworks of Planck and Carathéodory, the underlying hypotheses that lead to the traditional view of the Third Law as an independent postulate are examined. It is argued that the historical rejection of Nernst’s proof---motivated by Einstein’s insistence on the practical non-performability of cycles at absolute zero---overlooks the fact that a universal Second Law already precludes such cycles, rendering an independent Third Law an unnecessary complexity. Ultimately, the Nernst theorem is shown to be an essential consistency regulator rather than an independent physical first principle.

%% file: mainCount.tex
8813

%% file: keywords.tex

%% file: manuscript.tex
\section{Introduction}
\label{sec:introduction}
Thermodynamics is a branch of physics deeply rooted in undisputed empirical evidence. However, there remains a diverse range of perspectives on how this evidence should be formalized. Classical presentations of thermodynamics---often framing the field around the ``four laws that drive the universe''\citep{Atkins2007}---were pioneered by \citet{Clausius1854} and \citet{thomson-1853} and later systematized in the influential textbooks by \citet{Poincare1892} and \citet{planck-1897}. In contrast, more axiomatic approaches, such as those by \citet{Caratheodory1909}, \citet{Landsberg1956}, \citet{Hatsopoulos1965}, \citet{callen-85}, \citet{Lieb1999}, \citet{gyftopoulos-05}, \citet{Lavis2025}, and \citet{Tasaki2026} prioritize a systematic, logically consistent development of the formalism, often to expose and resolve the cyclic arguments latent in classical treatments.

While energy conservation (the First Law) is a cornerstone of every branch of physics, and the principle of entropy increase (the Second Law) holds what \citet{eddington-1915} called the ``supreme position among the laws of physics,'' the status of the remaining two laws is discussed. Thermal equilibrium and thermometry (the so-called Zeroth law) is not a first principle but a consequence of the principle of entropy increase. The Third Law, referring to properties at the boundary of accessible states, is often excluded from the formal exposition of the theory,\citep{Uffink2007} and ``as part of the fundamental theory of thermodynamics remains contentious,''\citep{Lavis2025}. The Third Law is also subject of considerable debate, confusion, and ongoing clarification.\citep{Falk1959,Bazarov1971,Yan1988,Landsberg1989,Oppenheim1989,Wheeler1991,Blau1996,landsberg-ajp-97,mafe-ajp-98,roseinnes-ajp-99,Belgiorno2003,Martin-Olalla2003b,Kox2006,Klimenko2012,Beretta2015e,Masanes2017,Uffink2017,Su2022,Martin-Olalla2024f}

Broadly defined, the Third Law characterizes two general properties of systems at ``the boundary of accessible states''---the neighborhood of absolute zero: (i) the vanishing of specific heats as temperature vanishes, and (ii) the vanishing of the isothermal variation of entropy as temperature vanishes, known as Nernst's theorem or Nernst heat theorem.\citep{Nernst1924} These two general properties were synthesized in Planck's celebrated statement: ``as the temperature diminishes indefinitely the entropy of a chemical homogeneous body of finite density approaches indefinitely near to a definite value, which is independent of pressure, the state of aggregation and of the special chemical modification'' \citep{Planck1921}.

From a pure thermodynamic point of view, the vanishing of the specific heats ---and therefore the definiteness of the entropy at the boundary of accessible states--- is not a first principle but a straight consequence of thermal stability at $T=0$,\citep{Bazarov1971,Martin-Olalla2025d} which is supported by robust empirical evidence\citep{Nernst1912}, confirmed by microscopic quantuum models.\citep{Einstein1907,Debye1912, Sommerfeld1928}  Consequently, in this study the Third Law strictly refers to the Nernst's theorem, which is supported by indirect evidence ---namely the unattainability of the absolute zero isotherm--- and by direct observations of the vanishing of $\Delta S$ in chemical reactions, including phase transitions, as well as the vanishing of expansion coefficients.\citep{Nernst1906,Young1975} 

The foundational controversy regarding Nernst's theorem originates in the early proof by contradiction proposed by \citet{Nernst1912}. The unattainability of the absolute zero isotherm was deduced from the premise that a finite Carnot engine operating at $T=0$ would negate Planck's statement of the Second Law.

Einstein famously rebutted this by noting that such a Carnot engine cannot be realized in practice, regardless of the existence of a $T=0$ reservoir \citep{Einstein-1913, epstein-1937, Boas1960, kestin-1968ii}. Recently a direct proof of the theorem was presented, based on the idea that a Carnot engine must be able to operate at $T=0$, if only to measure $T=0$.\citep{Martin-Olalla2025c} This debate concerns the very limits of thermodynamic reasoning; for a comprehensive review of the historical stakes, see \citet{Kox2006}.

The uniqueness and inherent structural weakness of the Third Law can be identified directly in its formulation. Unlike the First and Second Laws, whose statements are sufficiently general to remain independent of specific thermodynamic coordinates, the statement of the Third Law explicitly requires an a priori reference to temperature and entropy. Consequently, it is impossible to evaluate its structural position within the framework of thermodynamics without a precise, prior definition of these properties. By extension, the mere existence of a Third Law implies the existence ---if only theoretical--- of substances that satisfy the First and Second Laws yet violate Nernst's theorem. Such a substance lies at the core of Einstein's historic rebuttal and may appropriately be termed an \emph{Einsteinian substance}. Crucially, the classical ideal gas does not fall in this category as it simultaneously negate the vanishing of the specific heats. The purpose of this work is threefold: first, to identify the series of operational anomalies this substance would inflict upon the Second Law; second, to clarify the underlying hypotheses that render the Second Law a seemingly ``incomplete'' principle requiring an independent Third Law to dictate an evolution criterion across the full temperature domain; and third, to identify the alternative hypotheses that allow the Second Law to remain self-sufficient. The following analysis relies strictly on macroscopic thermodynamic arguments, intentionally excluding microscopic or statistical descriptions regarding the limit $T=0$. By focusing on the formal architecture of the theory, this study unveils that the Third Law is not an arbitrary addendum, but a necessary requirement for logical consistency within the Second Law itself.

The article is structured as follows: Section~\ref{sec:form-struct-prer} establishes the formal architecture and prerequisites of the theory, recalling the axiomatic foundations of the First and Second Laws along with the core concepts essential for the subsequent analysis. Section~\ref{sec:statement-problem} revisits the historical argument between Einstein and Nernst. Section~\ref{sec:why-third-law} and its subsections examine the issue from the perspectives of physics and epistemology, linking the Third Law to fundamental processes, equations of state, and the formal axiomatic structure derived inherently from the Second Law. Table~\ref{tab:global} shows a summary of the forthcoming discussion.

\begin{table*}
  \centering
  \begin{tabular}{p{6cm}lp{2cm}p{2cm}}
    \toprule
    \textbf{Foundational Issue}&\textbf{See}&\textbf{Nernstanian}&\textbf{Einsteinian}\\
    \midrule
    Why are finite Carnot engines at $T=0$ impossible?&\S\ref{sec:statement-problem}&Mandated by the Second Law&\mbox{\emph{Ad hoc}/} practical\\\addlinespace
    Nature of the limit as $T\to0^+$&\S\ref{sec:plancks-post-second}&Continuous&Singular\\\addlinespace
    Does the Second Law provides a universal criterion for evolution?&\S\ref{sec:plancks-stat-second}&Yes&No\\\addlinespace
    Is a quasi-static adiabatic process reversible?&\S\ref{sec:ideal-adiab-proc}&Yes&No\\\addlinespace
    Does a succession of equilibrium states constitute a valid process?&\S\ref{sec:entropy-temperature}&Yes&No\\\addlinespace
    Can absolute zero be determined by a consistent null measurement?&\S\ref{sec:entropy-temperature}&Yes&No\\\addlinespace
    Is entropy consistently associated with $Q/T$ in reversible isotherms?&\S\ref{sec:entropy-temperature}&Yes&No\\\addlinespace
    Is the boundary of accessible states related to thermophysical properties?&\S\ref{sec:equations-state}&Yes&No\\\addlinespace
    Is supported by evidence?&&Yes&No\\\addlinespace
    \bottomrule
  \end{tabular}
  \caption{Comparison of the axiomatic consistency between the proposed Nernstian framework (based on Planck’s statement, see Figure~\ref{fig:master}A) and the Einsteinian rebuttal,\citep{Einstein-1913,epstein-1937,Boas1960} where Planck's statement is maintained but Nernst's theorem is rejected, see Figure~\ref{fig:master}B. Note that the foundational inconsistencies (havocs) associated with the Einsteinian framework are solved when the Nernst's theorem is assumed as an independent, additional law.}
  \label{tab:global}
\end{table*}

\section{Formal structure and prerequisites}
\label{sec:form-struct-prer}

The subject of study in thermodynamics is the ``system,'' which refers to any macroscopic, homogeneous, and isotropic body. Specifically, a system is a region of space upon which we focus our attention and which contains matter or energy. The analysis focuses strictly on equilibrium states, attained spontaneously by isolated systems, a principle occasionally referred to as the ``minus first'' law of thermodynamics \citep{Brown2001}.

The operational definition of a system establishes its fundamental deformation variables: volume (associated with the spatial boundary), internal energy, and the amount of matter. In the following description, the amount of matter remains fixed, meaning the system is closed and composition variables play no active role. The role of volume can be subsumed by other generalized mechanical variables $X$, such as magnetization, polarization, or elongation, depending on the physical characteristics of the system. This mechanical variable is accompanied by a conjugate pair $Y$---such as pressure, magnetic field, electric field, or tension---which characterizes a generalized force field that can be externally controlled by an experimentalist. Thermodynamics fundamentally concerns itself with transitions (processes) between these equilibrium states initiated by external actions.

A primary topic of foundational interest is the behavior of an adiabatic enclosure, defined by the condition that only mechanical actions on the system are permitted, meaning energy can only be transferred in the form of work \citep{gyftopoulos-05,Lavis2025}. Within such an enclosure, the system under study is the sole thermodynamic actor in the universe, analogous to a free particle in classical mechanics, accompanied only by suitable external mechanical devices that execute work.

The following adiabatic processes are of central relevance to this study: 
(i) a free expansion, where no external constraining force is exerted and no work is exchanged, meaning the energy of the system remains unaltered ($dU=0$) while its volume increases ($dV>0$); 
(ii) a Joule paddle-wheel experiment, where a descending weight supplies gravitational potential energy to the system via a spinning paddle, causing the energy to increase ($dU>0$) at constant volume ($dV=0$); and 
(iii) an ideal adiabatic process, where the system expands via incremental adjustments in the force exerted upon an unrestrictive wall. 
If this latter process is sequenced, a finite change in volume is coupled with a finite change in external forces. If this sequence of small force adjustments is dynamically reversed, the external force recovers its initial value, and the final volume approaches its initial value asymptotically as the step sizes decrease indefinitely. Following \citet{Caratheodory1909}, the formal development of thermodynamic theory relies fundamentally upon the axiom that such an ideal, quasi-static adiabatic process is reversible. This process is governed by the differential relation $dU + pdV = 0$, where $p$ is the internal pressure. Because the slope $(\partial U/\partial V)_\Gamma = -p$ is negative, energy decreases with expanding volume during an ideal adiabatic process. Furthermore, since expansion can only be sustained by reducing the magnitude of this negative slope, internal energy must be a convex function of volume along this path. 

Crucially, any two equilibrium states can be connected by an adiabatic path consisting, at most, of an ideal process (iii) combined with either a free expansion (i) or a paddle-wheel experiment (ii). If two states are linked by process (iii) alone, the adiabatic connection is bidirectional. Conversely, if process (i) or (ii) is required, the adiabatic link becomes strictly unidirectional because neither free expansion nor paddle-wheel dissipation can be adiabatically reversed: either state 1 is adiabatically accessible from state 2, or vice versa, exclusively. Within this framework, adiabatic accessibility establishes a total ordering of equilibrium states. State 1 is adiabatically accessible from state 2 if and only if a state function, named entropy, satisfies $S_1 \geq S_2$. However, an adiabatic enclosure alone can neither provide a metric for entropy nor determine whether a given state is hot or cold.

\begin{figure*}
\centering
  \includegraphics{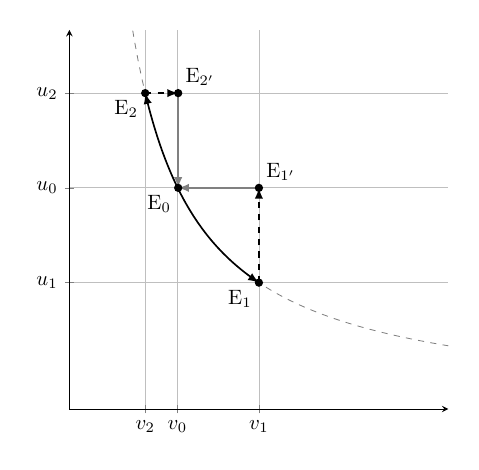}
  \caption{The ideal adiabatic process (thick line going through $\equilibrio_i$ ($i=0,1,2$)) and unary cycles. After the ideal adiabatic expansion $\equilibrio_0\to\equilibrio_1$, energy can be restored by a paddle-wheel experiment $\equilibrio_1\to\equilibrio_{1'}$ (broken line). Volume can only be restored by the non-adiabatic process $\equilibrio_{1'}\to\equilibrio_0$ (gray line) where heat is release at the expense of the compression work. Likewise, after the ideal compression $\equilibrio_0\to\equilibrio_2$, volume can be restored by a free expansion experiment $\equilibrio_2\to\equilibrio_{2'}$ (broken line). Energy can only be restored by the non-adiabatic process $\equilibrio_{2'}\to\equilibrio_0$ (gray line) where heat is released at constant volume  at the expense of the compression work in $\equilibrio_0\to\equilibrio_2$.}
  \label{fig:unary}
\end{figure*}

The second topic of interest concerns cyclic processes, where the final state of the working substance matches its initial state, thereby removing the net influence of the system's internal state changes. A purely adiabatic process can only form a closed cycle if it is entirely ideal. Otherwise, one deformation variable can be cycled, but the remaining variable cannot be restored unless the process is non-adiabatic (i.e., diathermal or diabatic). As an example, an ideal adiabatic expansion can be followed by a paddle-wheel experiment that restores the internal energy of the system, completing the energy cycle via the path $\equilibrio_0 \to \equilibrio_1 \to \equilibrio_{1'}$ illustrated in Figure~\ref{fig:unary}. To return the system to its initial state, its volume must be decreased, requiring negative work. Because the internal energy must remain net-unchanged over the complete cycle, this negative work must be rejected into an external system via a non-mechanical mode of energy exchange termed heat. As a result, mechanical work is irreversibly dissipated into heat. Conversely, an ideal adiabatic compression can be followed by a free expansion that restores the volume ($\equilibrio_0 \to \equilibrio_2 \to \equilibrio_{2'}$). To subsequently restore the initial internal energy, the system must release energy as heat, as compression work is unavailable. 

The above cycles are unary, meaning they require only a single thermal interaction with an external reservoir to complete the loop. Empirical evidence persistently demonstrates that unary cycles inevitably result in the net dissipation of work into heat and are strictly irreversible. The Second Law of thermodynamics elevates these physical observations to a primary axiom. In Planck's classic formulation: ``it is impossible to construct an engine which will work in a complete cycle, and produce no effect except the raising of a weight and the cooling of a heat-reservoir'' \citep{planck-1897}.\footnote{The term ``reservoir'' refers to an idealized thermal system of arbitrarily large heat capacity. It can be substituted by ``a body'' without loss of generality; the underlying physical constraint is the directional restriction of the energy exchange regardless of the absolute scale of the thermal transfer.} This statement is a masterpiece of conceptual conciseness; it formally establishes the reality of irreversibility and operationally defines heat as a mode of energy transfer that, unlike work, is bound by a fundamental directional asymmetry, bypassing the need for complex caloric or microscopic characterizations of thermal interactions \citep{Beretta2015c}.

The Second Law formalizes the early observation by \citet{carnot-1824} that motive power is derived from the transport of ``caloric'' from a hot to a cold body, alike the motive power extracted for a watermill. Indeed, the most direct consequence of this statement is that raising a weight via a cyclic process necessitates two distinct heat exchanges: (i) with a ``boiler'', which is cooled (rejects heat, mandated by the First Law), and (ii) with a ``cooler'', which must be heated (accepts heat, mandated by the Second Law). This is a binary engine, namely a Carnot engine. The First Law mandates that:
\begin{equation}
  \label{eq:8}
  \uQ_b+\uQ_c=\uW,
\end{equation}
where $\uQ_b$ is the heat exchanged with the boiler, and $\uQ_c$ the heat exchanged by the cooler.

The Second Law ($\mathrm{SL}$) leads to three material implications:
\begin{align}
  \label{eq:2}\mathrm{SL}\Longrightarrow& (\uW>0\Rightarrow\uQ_c<0),\\
  \label{eq:7}\mathrm{SL}\Longrightarrow &(\uW\leq0\Leftarrow\uQ_c=0),\\
  \label{eq:7b}  \mathrm{SL}\Longrightarrow &(\uW<0\Leftarrow\uQ_c>0),
\end{align}
where $\uW$ is the work performed. 

 Relation~(\ref{eq:2}) formalizes the statement itself: if a weight is raised, a third effect (the heating of a cooler) must occur. Relation~(\ref{eq:7}) describes the dissipation of work into heat, while relation~(\ref{eq:7b}) characterizes refrigeration: if a cooler is cooled, an amount of work must be performed (a weight must be lowered).

 The Second Law introduces the temperature as a physical observable. The point to note is that the necessity of a boiler and a cooler to convert heat into work is not a mere cardinal necessity of any two elements, but they must be necessarily arranged in a certain way. In other words, the roles played by the boiler and the cooler cannot be exchanged. This aligns with the ``less than'' binary relation, which does not allow this exchange either.\citep{Ehrlich1981} This is the asymmetric property of a binary relation. In this context the Second Law puts forward that Carnot engines are irreflexive: a body cannot play the two roles, boiler and cooler, in the engine. The two properties  orders the set of thermodynamic systems. The temperature $T$ is the state variable that translates this order into the set of real number: the boiler and the cooler must sustain $T_c<T_b$ for some state variable.  In this scope temperature aligns with the capability of producing work with thermal engines. It also aligns with the common perception that heat flows in a predetermined sense: from hot to cold.

 In summary, while the concept of temperature historically predates the Second Law\citep{Uffink2007} and many presentations of thermodynamics rush to define thermal contact, the thermal equilibrium and the [empirical] temperature prior to addressing the conversion of heat into work, temperature is not logically anterior to the Second Law, but one of its consequences. It must also be noted that while the Second Law was empirically deduced from synthesis of observations at ``room temperatures'', its bare statement is, \emph{prima facie}, more general and unspecific.

Carnot's theorem establishes reversible engines as the theoretical limit of efficiency.\citep{carnot-1824} Consequently, reversible exchanges in reversible Carnot engines are governed by the properties of nature and not by experimentalist's will or capacity. Hereafter, quantities exchanged in a reversible process are denoted by sans-serif letters, such as $\Wrev$ or $\Qrev$.

Reversible heat exchanges can be linked to the thermophysical properties of the working substance and the reservoirs. First, reversible engines establish a universal metric for temperature, the Carnot's temperature $\ut$, defined by the ratio:
\begin{equation}
  \label{eq:3}
  \frac{\ut_c}{\ut_b}:=-\frac{\Qrev_c}{\Qrev_b},
\end{equation}
where energy conservation~(\ref{eq:8}) ensures that $-\Qrev_c<\Qrev_b$ and, therefore, $\ut_c<\ut_b$, a condition sufficient to characterize the ordering isomorphism described above. This fundamental relation was noted first by \citet{thomson-1848}. In irreversible binary engines the equality in~(\ref{eq:3}) becomes a strict ``less than'' inequality.

Second, reversible engines provide a metric for the entropy, known as Clausius' entropy defined differentially as:
\begin{equation}
  \label{eq:4}
  dS:=\frac{\Qrev}{\ut}.
\end{equation}
Entropy is conserved in reversible Carnot engines as a direct algebraic consqeuece of~\eqref{eq:3}, and, by extension, it remains a state function of any arbitrary reversible cyclic process. Irreversible cycles transform the equality into a ``greater than'' strict inequality..

The definition~\eqref{eq:4} is the culmination of Clausius' theorem. The standard proof of this theorem considers an arbritary cycle involving a series of localized heat exchanges $Q_i$ occurring at corresponding temperatures $\ut_i$. Each local reservoir at $\ut_i$ is coupled to an auxiliary reversible Carnot engine that balances the exchange such that $Q_i+\Qrev_i=0$. The second reservoir of each auxiliary reversible Carnot engine operates at a fixed, common reference temperature $\ut_o$, independent of $T_i$. The composite cycle, comprising the original cycle and the network of auxiliary reversible Carnot engines constitutes, by design, an unary cycle whose only net heat exchange $\Qrev_o$ occurs at $\ut_o$. The Second Law mandates that $\Qrev_o \leq 0$, where the equality holds strictly for entirely reversible processes. Crucially, this implies that the formal definition of Clausius' entropy as a state variable is supported exclusively by reversible Carnot engines, Carnot's theorem, and, consequently, the Carnot's temperature. 

Combining \eqref{eq:3} and~\eqref{eq:4} with energy conservation (\ref{eq:8}), work in a reversible Carnot engine can be expressed as:
\begin{equation}
  \label{eq:5}
  \Wrev=(\ut_b-\ut_c)\cdot\Delta S,
\end{equation}
where $\Delta S=\Qrev_b/\ut_b=-\Qrev_c/\ut_c$ represents the finite entropy flux from the boiler to the cooler. Equation~(\ref{eq:5}) formally expresses the early insight by \citet{carnot-1824} that work is produced by the transport of an agent from hot body to cold one; here entropy takes on the operational role originally envisioned for caloric.

The above framework describes Carnot engines as binary engines where only two heat exchanges occur. This suffices to underpin the whole physics behind the problem. Notwithstanding this, Carnot engines can be described thoroughly by thermodynamics cycles where state variables change in the appropriate manner. Usually, the Carnot cycle is a four-stroke cycle composed of two isotherms where heat is absorbed (boiler) and rejected (cooler) and two ideal adiabatic processes, where entropy remains constant. In a $ST$ chart this cycle looks like a rectangle, irrespective of the properties of the working substance, and equation~\eqref{eq:5} is just the area of the rectangle.

However fundamental this setting can be, it is not the only way of achieving a binary cycle because the true requisite for the engine to be binary is that the two paths connecting either isotherms differ by a constant shift of entropy equal to the entropy exchanged at the boiler.\citep{Martin-Olalla2003b} With Equation~\eqref{eq:4} insensitive to a constant shift of the entropy, the net heat exchanged along these two paths cancel to each other for every temperature other than $T_b$ and $T_c$, and the cycle is governed by the same balance equations as a regular two-isotherm, two-isentropic cycle. In a $ST$ chart, the new cycle is rendered as a simple quadrilateral with two sides upright, representing the two isotherm processes. The area remains $\Delta T\Delta S$.

The above framework leaves open a significant question: if temperature is derived from the Second Law, and the Second Law forbids an engine with $Q_c=0$, how can $T=0$ be formalized in view of \eqref{eq:3}? Historically, this question was left unaddressed by \citet{thomson-1848}, who recognized the absoluteness of the temperature defined by Carnot's theorem but made no remark on the zero of the ``scale.'' Formally, it is often assumed that Carnot's theorem \eqref{eq:3} does not apply at $T=0$, meaning that the domain of Carnot's temperature is strictly $T>0$ and that ``the kelvin scale of temperature is not defined at $T=0$.'' \citep{liboff-physicsessays-94,Chen2026} However, from a conceptual perspective, a physical observable like temperature requires a meaningful characterization of its null value. Because every physical and formal aspect regarding temperature and entropyc---including the determination of their legitimate domains--- must be derived strictly from the intrinsic properties of reversible Carnot engines, it is uniquely within the scope of the Second Law that this boundary problem must be addressed. The following section will discuss this issue thoroughly, but a brief answer can already be outlined: strictly speaking, the Second Law does not forbid an engine operating with $Q_c=0$; rather, it forbids the raising of a weight ---that is, the production of a positive net work, $W>0$--- when $Q_c=0$.

\section{Why is a third law of thermodynamics needed?}
\label{sec:statement-problem}

\begin{figure*}
  \includegraphics[width=\textwidth]{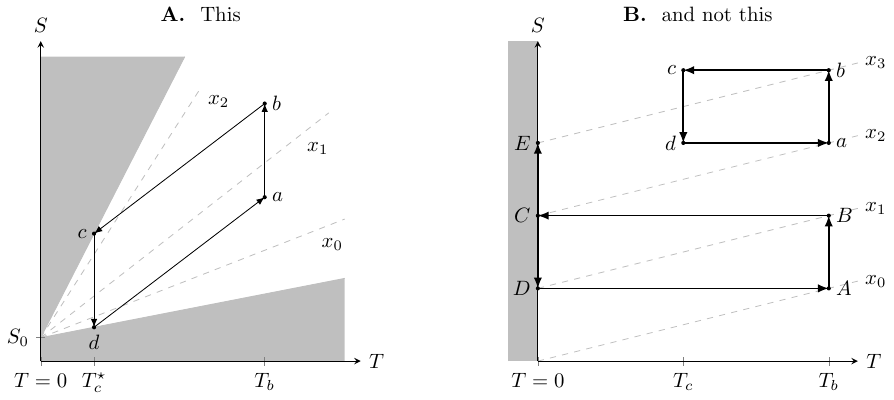}
  \caption{Comparison between observed (A, left) and hypothetical (B, right) thermodynamic behaviors as $T \to 0$.\citep{Masanes2017} Panel A shows the standard behavior consistent with the Nernst theorem (Third Law) while panel B depicts a substance violating the Nernst theorem. Gray shaded regions indicate inaccessible states within the domain of $x$, a mechanical parameter.  In panel A, $abcda$ is a generalized Carnot engine consisting in two isothermal processes and two processes that differ in a constant shift of entropy. Note that the area enclose remains $\Delta T\time\Delta S$ as in two-isotherm, two-adiabat Carnot engines. The temperature $\ut_c^\star$ is the lowest temperature at which the engine can operate for the given working substance and the given $\Delta S=S_b-S_a$.\citep{Martin-Olalla2003b} In panel B the cycle $abcda$ is a Carnot engine operating between non-zero temperatures $\ut_b$ and $\ut_c$; the cycle $ABCDA$ is a Carnot engine operating at $T=0$.}
  \label{fig:master}
\end{figure*}

Figure~\ref{fig:master} illustrates the foundations of the classical debate regarding the interplay between the Second and Third Laws of thermodynamics. Panel A displays the phase space of a substance adhering to both the vanishing of specific heat and the Nernst theorem: all iso-$x$ lines converge to a unique, finite entropy value $S_0$ at $T=0$, independent of the mechanical variable $x$. Known homogeneous systems consistently follow this behavior, with the caveat that in pure hydrostatic systems (gases), the volume domain is physically constrained by the container dimensions \citep{landau-lifshitz-1968}.

Conversely, panel B represents a hypothetical substance where the specific heat vanishes as $T \to 0$, yet the Nernst heat theorem is violated. This behavior is characterized by a manifold of accessible entropy values at absolute zero; notably, the iso-$x$ lines (where $x$ denotes a mechanical parameter such as volume or magnetization) maintain finite, non-zero positive slopes, ensuring that the specific heat vanishes despite the existence of an entropy residual. Such a substance is purely pedagogical and has never been observed in nature. Fundamentally, it characterizes a world governed solely by the First and Second Laws; for reasons detailed below, it may appropriately be termed an \emph{Einsteinian substance}. The remainder of this paper is dedicated to dissecting the anomalous properties, operational boundaries, and theoretical caveats that emerge when this hypothetical substance is subjected to the absolute limits of the Carnot cycle.

In both panels, gray shaded regions indicate inaccessible states—pairs of $(T,S)$ that do not represent equilibrium states for the system within the defined domain of $x$. In figure~\ref{fig:master} $T=0$ is set as a boundary that separates accessible states from unaccessible states by virtue of~\eqref{eq:3} with $Q_c<0$ (the cooler is heated), see later to section~\ref{sec:equations-state}.

The classical discourse can be succinctly summarized as follows: \citet{Nernst1912} argued that because a Carnot engine $ABCDA$ operating at $T=0$ would allow the total conversion of heat into work, it would fundamentally negate the Second Law Consequently, Nernst maintained that $T=0$ must be inaccessible, rendering the convergence in panel A a theoretical necessity, which aligned with his empirical observations.

However, Einstein countered that the $ABCDA$ cycle is physically impossible for independent reasons: in $CD$ any infinitesimal imperfection would displace the system from $T=0$, meaning the engine cannot be realized ``in practice'' and thus cannot formally challenge the Second Law \citep{Einstein-1913}. \citet{epstein-1937} further noted that ``no experimental direction is given (or, indeed, possible) to insure that a compression, starting from point B [$C$ in figure~\ref{fig:master}] should cause the system to move in the curve BC [$CD$ in figure~\ref{fig:master}] and not in BA [$CB$ in figure~\ref{fig:master}].'' Similarly, \citet{Boas1960} observed that ``a Carnot cycle with a $T=0$ reservoir cannot in practice be performed, regardless of the existence or not of the $T=0$ reservoir''. From this perspective, the behavior in panel A is viewed not as a deduction from the Second Law, but as a distinct, independent empirical observation.

\citeauthor{Nernst1912} remained unconvinced, pointing out that Einstein’s dismissal of reversible transformations at $T=0$ would ``put into question numerous other thermodynamic theorems'' \citep{Simon1951,Kox2006}. While subsequent sections will detail the formal thermodynamic inconsistencies introduced by Einstein's notion, it is instructive to consider a remarkable observation made by \citeauthor{Nernst1924}. If a substance such as the one in Figure~\ref{fig:master}B were to exist, an experimentalist could perform an ideal adiabatic process where the mechanical parameter changes monotonically ($x_1 \to x_2 \to x_3$, path $B \to C \to E$ in Figure~\ref{fig:master}B). \citeauthor{Nernst1924} noted that along $BC$, the conjugate mechanical parameter $y$ would change according to the adiabatic susceptibility $(\partial y/\partial x)_s$, while along $BE$, it would follow the isothermal susceptibility $(\partial y/\partial x)_t$, after the boundary is reached. This would result in a spontaneous discontinuity in the system's response despite a continuous experimental action ---such as the slow rotation of a control knob. To illustrate this point, \citeauthor{Nernst1924} utilized the example of a gas expanding according to the adiabatic law $pV^\gamma = \text{const.}$ ($\gamma > 1$) followed by the isothermal law $pV = \text{const.}$, resulting in a perceptible and physically unappealing change in the slope of the expansion path. Section~\ref{sec:ideal-adiab-proc} and section~\ref{sec:equations-state} will refine Nernst's point of view. Simple, intuitive and significant as this remark is it has never been countered thoroughly in literature; we may say it has just been ignored.

At that time, the vanishing of the specific heats at absolute zero, another empirical observation noted by \citet{Nernst1912}, also caused a major shock in the scientific community because classical physics could not seemingly provide a reason. \citet{Einstein1907} and \citet{Debye1912} provided a theoretical explanation for this observation based on the emerging quantum mechanics. While the vanishing of the variation of entropy and the vanishing of the specific heats are fundamentally independent from a theoretical point of view\footnote{One can design a system that fulfills the vanishing of the specific heats and, alternatively, obeys or not Nernst theorem, as in figure~\ref{fig:master}. Also one can design a system that fulfills Nernst theorem and does not comply with the vanishing of the specific heat, for instance adding a divergent contribution like $\log T$ to figure~\ref{fig:master}A.} they shared some common grounds from an empirical perspective. Nernst himself attempted to link his theorem to quantuum properties\citep{Simon1951,Kox2006} while Einstein was probably reluctant to admit that a classical theory could provide a consistent explanation for a macroscopic observation at $T=0$. Consequently, Einstein leaned towards the necessity of a new, independent law. 

\section{Why is a third law of thermodynamics NOT needed?}
\label{sec:why-third-law}

\citet{Landsberg1956} contends that the crux of the controversy lies in the status of boundary points (equilibrium states at $T=0$) relative to an open set of points (regular equilibrium states at non-zero temperature). He notes: ``In mathematics the status of such points is often the subject of a separate assumption. Nernst assumed that, in the absence of information to the contrary, the points could be regarded as belonging to the set. Einstein suggested that they should not belong to the set. From the axiomatic point of view, however, both views represent an additional postulate.''

While this interpretation holds in certain abstract mathematical contexts, the physical challenge is to validate the consequences of the framework originating from the Second Law and determine if these consequences align with empirical evidence---particularly in the limit as $T \to 0$. This section demonstrates the various ways in which the Second Law implicitly predicts the Nernst heat theorem, and the various ways in which the framework originated by Einstein's rebuttal fails to address foundational issues. 
\subsection{The continuity the Second Law at the boundary}
\label{sec:plancks-post-second}

A consistent physical theory requires that the consequences of the Second Law do not merely apply to the interior of the state space, but remain continuous and well-behaved as the boundary of the state space is approached. This section is devoted to show that this idea derives Nernst's theorem from Second Law.

For reversible engines, relation~(\ref{eq:7}) becomes more stringent: the Second Law mandates that if a reversible engine rejects no heat into the cooler, the engine must produce no net work:
\begin{equation}
  \label{eq:1}
  \mathrm{SL}\Longrightarrow (\Wrev=0\Leftarrow\Qrev_c=0).
\end{equation}
This relation must hold not only as a point assigment, but also as a limit:
\begin{equation}
  \label{eq:9}
  \mathrm{SL}\Longrightarrow (\Wrev\to0^+\Leftarrow\Qrev_c\to0^-),  
\end{equation}
because reversible exchanges are linked to the thermophysical properties of the working fluid and the reservoirs, see~(\ref{eq:3}) and~(\ref{eq:5}). In a consistent physical theory, these properties must be continuous and well-behaved as the boundary of the state space is approached; consequently the behaviour of $\Qrev_c$ must be governed by the same continuity.

Relation~(\ref{eq:9}) is the Nernst's theorem illustrated in Figure~\ref{fig:master}A with $\Qrev_c$ linked to $\ut_c$ as per~(\ref{eq:3}) and $\Wrev$ linked to $\Delta S$, as per~(\ref{eq:5}). Nernst's theorem is proven with the following chain of material implications:
\begin{equation}
  \label{eq:6}
  \ut_c\to0^+\Rightarrow\Qrev_c\to0^-\Rightarrow\Wrev\to0^+\Rightarrow\Delta S\to0^+.
\end{equation}
In the first implication, the case $\Qrev_b\to\infty$ is excluded by the principle of energy conservation.\citep{Heidrich2016} This proof of consistency follows directly from the Second Law without the need to hypothesize the negation of the theorem to seek a contradiction, as was required in Nernst's original proof.

Relation~(\ref{eq:9}) can be understood as follows: the Second Law requires that a non-zero heat $\Qrev_c$ is rejected if work is produced; in a logical context, the non-vanishing character of $\Qrev_c$ ---how much different from zero $\Qrev_c$ needs to be--- is determined by the laws of nature, not the experimentalist. In Figure~\ref{fig:master}A, if unit heat is extracted at a unit temperature, the thermophysical properties of the working fluid establish a lower bound for $\Qrev_c$, dictated by the limit temperature $\ut_c^\star$, which is determined by the amount of entropy exchanged with the boiler.

In Figure~\ref{fig:master}B, the Einsteinian perspective suggests that an experimentalist could theoretically set $\Qrev_c$ arbitrarily close to zero while maintaining finite net work by exploiting specific fluid properties. However, this introduces a fundamental continuity problem: if a reversible Carnot engine is strictly forbidden at the boundary $T=0$, a consistent physical theory requires that both the engine's capacity for work and the working substance's capacity to transport entropy must vanish continuously as that limit is approached. To suggest otherwise is to treat the Second Law not as a robust, universal principle of nature, but as a fragile mathematical point-discontinuity that an experimentalist can circumvent asymptotically. In the classic analogy of the watermill, such a framework implies that the mill would conveniently stop spinning at the exact moment its power output reaches its absolute maximum.

This line of thought may be summarized as: if an engine provides some finite $\uW$, then it must deliver some finite $\uQ_c$ to a cooler. This requirement ---a coupling between $\uW$ and $\uQ_c$--- is implicit in the statement of the Second Law and, arguably, does not require a separate formalization \citep{Martin-Olalla2003b}. It must be noted that the statement of the Second Law does not make any reference to state properties of systems. It only refers to work and heat. Consequently, it cannot explicitly contain any reference to the continuity issue here described. This can only arise from the framework that is logically derived from the law. Finally, we are not supporting a mere exchange of two distinct principles ---continuity in exchange of the Nernst theorem---. Continuity is a general prerequisite embedded in physical properties, not a thermodynamic-specific requirement.

\subsection{The ontological consistency of the Second Law}
\label{sec:plancks-stat-second}

Notwithstanding the derivation in Section~\ref{sec:plancks-post-second}, this section shows an alternative line of reasoning that directly links the Second Law to the Nernst heat theorem.

In the classical rebuttal of \citeauthor{Nernst1912}'s proof, much is made of the ``point of logic'' connecting Nernst's contradiction to the hypothetical experiments that would be feasible if the behavior in Figure~\ref{fig:master}B were possible.\citep{epstein-1937, Boas1960,Bazarov1971} However, these remarks may ultimately be irrelevant to the core theoretical discussion.

The First and Second Laws constitute the axioms of a framework that yields both the concept and metric of temperature (\ref{eq:3}), and the concept and metric of entropy (\ref{eq:4}), see section~\ref{sec:form-struct-prer}.  This framework is built upon the postulate that the Second Law formalizes a criterion of evolution for discriminating whether a process is possible. This criterion is typically expressed via the law of entropy increase and is universally accepted as robust at non-zero temperatures.

At $T=0$, Einstein's rebuttal effectively maintains that this criterion is no longer operative; he suggests that the Second Law is not the arbiter for the $ABCDA$ cycle, but rather that practical, extrinsic limitations render the process impossible.  While the Third Law, as an independent postulate, restores this criterion at the boundary, one must ask: is this the most parsimonious way to describe the laws of nature? The whole discussion in this present study is defining the consequences for the framework of thermodynamics that this ``practical'' criterion leaves behind, a point early noted by Nernst himself.

It is crucial to note that Planck's statement of the Second Law did not introduce a previously unknown property---it was already well understood that heat cannot be fully converted into work, just as it was known that heat does not flow spontaneously from cold to hot. Instead, \citet{planck-1897} elevated a known physical observation to a first principle from which all subsequent consequences are derived. As long as these consequences do not contradict empirical evidence, the statement prevails.

A commitment to Planck's statement as a universal criterion for discriminating whether a process is possible yields the Nernst theorem---for instance, through Nernst's original proof by contradiction---which aligns perfectly with experimental evidence. 

Ultimately, Planck's statement is sufficient to explain why substances of the kind depicted in panel B are not observed in nature. Is it relevant that an alternative line of thought may provide a different explanation on how experimentalists would handle finite Carnot's engines in the neighborhood of absolute zero, given that a definitive experimental test to distinguish between these two viewpoints remains fundamentally unavailable?

\subsection{Ideal isotherms at $T=0$}
\label{sec:entropy-temperature}

Much of the early 20th-century discourse regarding the boundary of accessible states centered on the characterization of an ideal (reversible) isothermal process at $T=0$. This discussion was largely dominated by the paradoxes inherent in the behavior depicted in Figure~\ref{fig:master}B. \citeauthor{Nernst1924} maintained that if, \emph{ex hypothesi}, $T=0$ is reached via a process occurring in ``finite dimensions'' (i.e., non-asymptotically), such as path $BC$ in Figure~\ref{fig:master}B, then the ``compression'' $CD$ must be possible; at this limit, internal energy would coincide with free energy, the thermodynamic potential driving an ideal isotherm. Conversely, \citet{Einstein-1913} and \citet{epstein-1937} argued that any isotherm at absolute zero must effectively be adiabatic, and that a compression would inevitably displace the system from $T=0$, ideally returning it to state $B$. Interestingly, this latter view relies heavily on empirical knowledge derived from substances that already satisfy the Nernst theorem, rather than those behaving like Figure~\ref{fig:master}B, which is empirically unknown.

In a sense, both Nernst's and Epstein's perspectives are simultaneously valid and flawed. The crux of the issue is that Figure~\ref{fig:master}B presents a bifurcation at point $C$: the experiment ``adiabatic compression from $x_2 \to x_1$'' offers two distinct paths: $CD$ (zero heat because $T=0$, as preferred by Nernst) and $CB$ (zero heat because $dS=0$, as preferred by Epstein). 

Beyond these historical ``mind games,'' ideal isotherms are formally required to establish the metrics of both temperature (\ref{eq:3}) and entropy (\ref{eq:4}). They are, therefore, essential to the definition of the $T=0$ axis itself. In this context, Equation~\eqref{eq:4} shows an obvious singular point at $T=0$, and this limit must be consistently analyzed.

For that to occur, as $T$ decreases indefinitely, $\Qrev$ must also decrease indefinitely in such way that their ratio remains finite, representing the isothermal entropy variation driven by a \emph{reversible} heat exchange at temperatures indefinitely low. In the limit, this must also account for the entropy variation associated with the \emph{reversible} exchange of a null heat at absolute zero. Consequently, this ratio can only be zero; because any non-zero value would lead to a negation of the Second Law as long as we still speak of a finite, non-zero, \emph{reversible} $\Qrev/\ut$ in the limit $\ut\to0$.

Paradoxically, denying the existence of reversible isotherms at $T=0$ (following Einstein's logic) while simultaneously allowing for a range of entropy values at $T=0$ (as in Figure~\ref{fig:master}B) creates a formal contradiction with the metric provided by~\eqref{eq:4}, which refers to isothermic reversible exchanges of heat: in Figure~\ref{fig:master}B, states $C$ and $D$ can only be reversibly connected via a non-isothermal path such as $CBAD$. This path certainly allows for the mathematical computation of the limit $\Qrev/T$ as $T\to0$, however, it fails to align with definition (\ref{eq:4}). In summary, for the Einstein's point of view, the metric~\eqref{eq:4} would only be valid for $T>0$; nonetheless, the framework would still be able to provide a value for the ``entropy at $T=0$''. Alternatively, if the $S$ in Nernst's theorem is understood strictly as Clausius' entropy~\eqref{eq:4}, then the theorem is proven by definition.

As noted in section~\ref{sec:form-struct-prer}, temperature characterizes the order relation associated with the conversion of heat into work as derived from the Second Law. As long as temperature is a state property and, therefore, a physical observable its zero value must be characterized by the Second Law through~\eqref{eq:3}. Equation~\eqref{eq:3} is the only one that provides a universal metric for temperature, therefore is the only one that can provide a meaningful, universal zero beyond schematic, specific descriptions like the temperature at which the volumen vanishes, which originated in the 18th the first evaluation of the absolute zero. In the Nernst-Einstein debate $T=0$ is seen as $p=0$.

Conversely, the historical evolution of thermodynamics and the ubiquitous reliance on empirical temperature (now formalized by the Zeroth Law) since the early eighteenth century have obscured a critical fact: in Nernst's theorem, temperature can only logically be that which arises from the Second Law. Consequently, $T=0$ is operationally defined by a reversible Carnot engine operating with $\Qrev_c=0$\citep{Martin-Olalla2025c}. This assignment is universal and leads directly to the Nernst heat theorem, as demonstrated in \eqref{eq:6}. In short, if the temperature $T$ in Nernst's formulation is understood strictly as Carnot's temperature \eqref{eq:3}, then the theorem is proven by definition. Otherwise, one would be forced to accept the contradiction that $T=0$ lacks a universal, meaningful physical characterization while simultaneously performing operational limits such as $T \to 0$.

The assignment $T=0$ requires a limit process that is necessarily consistent with Planck's statement of the Second Law.\citep{Martin-Olalla2026b} Figure~\ref{fig:master}A illustrates the generalized Carnot's cycle described in section~\ref{sec:form-struct-prer} $abcda$ consisting of two reversible isotherms ($ab$ and $cd$) and two processes ($bc$ and $da$) differing only by a shift in entropy. With the heat exchanges in $bc$ and $da$ canceling out, the cycle $abcda$ remains a binary engine. In $ab$, the experimentalist determines a finite $\Delta S$ taken from the boiler requiring a mechanical action $x_1\to x_2$. In $cd$, the full domain of the mechanical parameter is utilized to reject that same $\Delta S$ at the cooler's temperature, measured by the ratio $\Qrev_c/\Delta S$. Making $x_2$ closer to $x_1$, $\Delta S$ at the boiler will decrease and $T^\star_c$ will decrease as well. As $x_2$ approaches $x_1$ at $T_b$, $\Delta S$ approaches zero, in the limit the four-stroke engine degenerates into a round trip: a two-stroke engine with unit efficiency but no net heat or work, preserving the integrity of Planck's statement. In contrast, applying this to Figure~\ref{fig:master}B reveals the fundamental problem: at $T=0$, the cycle either negates the Second Law or becomes ill-defined, leaving absolute zero without a meaningful thermodynamic characterization.

The difference between these two constructions is fundamental: in the first case (Figure~\ref{fig:master}A), the operational property that allows the thermometer to function is entropy; in the second case (Figure~\ref{fig:master}B), it is the heat supplied to the cooler. While this distinction remains irrelevant as long as $T>0$, it matters profoundly at $T=0$. In a way, the conventional view that Carnot's theorem applies only for $T>0$ stems from \citet{thomson-1848}'s early perspective that heat exchanges in a Carnot engine provide a universal metric for temperature---a view inherited from \citet{carnot-1824}'s postulate that caloric is conserved in reversible engines. However, it is entropy, rather than heat, that explicitly fulfills these two prominent roles.

\subsection{Ideal adiabatic processes at $T=0$}
\label{sec:ideal-adiab-proc}

In section~\ref{sec:form-struct-prer} we introduced the concept of ideal adiabatic processes as one kind of process in which mechanical actions are small enough so that finite adiabatic transformations can be reverted and the initial state recovered. In short, ideal adiabatic processes are reversible.

In this description temperature plays no role. First, ideal adiabatic processes do not require the kind of comparison that is associated with temperature ---either thermal contact to allow the flow of heat, or a Carnot engine to test the production of work. Second, temperature is not required to describe the reversibility of the ideal adiabatic process. During this process, the system remains isolated, free from thermal interactions with any other system, thus behaving analogously to a free particle in classical mechanics. The experimentalist simply exerts a smooth mechanical action on the system, for instance, by turning a control knob. This action monotonically varies the coordinate $x$, causing the conjugate variable $y$ to adjust accordingly. Reversing the action would yield the exact reverse physical results as $y$ would also reverse. Consequently, the characterization of ideal adiabatic processes as reversible is, \emph{prima facie}, a valid assumption across the entire temperature domain.

In Figure~\ref{fig:master}B, $B\to C$ is performed by doing the kind of mechanical actions that are described for an ideal adiabatic process with the mechanical variable changing $x_1\to x_2$. The mechanical actions can be reverted $x_2\to x_1$, and the state of the system will follow $C\to B$, making the process reversible. Now, after $B\to C$ the experimentalist continues to proceed with the same mechanical actions $x_2\to x_3$ the state of the system would go, \emph{ex hipotesi}, towards $E$. If the mechanical actions are now reverted $x_3\to x_2\to x_1$ the state of the system will not attain $B$ because $E\to C$ is vetoed under Einstein's logic. Instead, the system will reach another state ---not shown in the figure--- resulting from the intersection of the entropy at $E$ with the $x_1$ line. An ideal adiabatic process will not be reversible.

Under Einstein's logic, we need a restrictive auxiliary assumption: that an ideal adiabatic process is reversible \emph{only if $T>0$}. This is theoretically problematic because (i) temperature is alien to the characterization of ideal adiabatic processes, as noted above; and (ii) one must introduce this restriction only to later formalize a third, independent first principle whose sole purpose is to remove the auxiliary assumption because if the Nernst's theorem is set forward then ideal adiabatic processes are reversible.

Instead, if the reversibility of ideal adiabatic processes is formalized as a general prerequisite, then the Nernst theorem is given. Following the logic of \citet{Caratheodory1909}, we may conclude that simple systems must, by definition, obey the Nernst theorem. Therefore, the persistent observation that finite-density, homogeneous substances abide by the theorem serves as empirical evidence that they can be treated as simple systems down to $T=0$.

\subsection{Negative temperatures and the boundary of accessible states}
\label{sec:equations-state}

This section reformulates the analysis presented in Section~\ref{sec:ideal-adiab-proc}, focusing on the criticism raised by \citeauthor{Nernst1924} regarding the process $B \to C \to E$ in Figure~\ref{fig:master}B, as introduced in Section~\ref{sec:statement-problem}.

Historically, the existence of an absolute zero temperature was deduced from the behavior of non-condensed phases, specifically, by extrapolating isobaric expansion experiments down to the limit $V\to 0$. In this view, negative volumes physically precluded the existence of negative temperatures. In the classic debate between \citeauthor{Nernst-1912} and \citeauthor{Einstein-1913}, $T=0$ is treated as such an empirical circumstance. Later, the domain boundary of equation~\eqref{eq:3} was used to support $T=0$ as the lowest attainable temperature within the framework of the Second Law. This latter perspective is fundamentally different, as the restriction is no longer merely empirical but emerges as a direct mathematical consequence of an axiomatic statement. Notwithstanding this, Carnot engines operating with negative temperature have been discussed ealier, in a framework compliant with the Third Law of thermodynamics, i.e. with the crossing point at $T\to\infty$.\citep{Landsberg1977}

\citeauthor{Nernst-1912}’s point of view is self-consistent in this regard. Recognizing the empirical difficulty of attaining absolute zero, he constructed a proof by contradiction that led to a substance whose thermophysical properties render $T=0$ a singular boundary: specifically, an accumulation point in the $ST$-plane (see Figure~\ref{fig:master}A). This makes $T=0$ unattainable.

Conversely, the Einsteinian point of view is less consistent because, for an Einsteinian substance (Figure~\ref{fig:master}B), $T=0$ is a temperature state as regular as any other. Nevertheless, \citeauthor{Nernst-1912} still identified the boundary shown in Figure~\ref{fig:master}B and, abhorring \emph{ex hypothesi} negative temperatures, formulated his argument by asserting that the path $B \to C \to E$ would suffer a discontinuity at $T=0$, see section~\ref{sec:ideal-adiab-proc}. During this process, the experimentalist exerts a smooth mechanical action on the system in the precise direction that the temperature decrease. It is always possible to choose such effect as long as $(\partial T/\partial x)_s$ is non-zero as in figure~\ref{fig:master}B.

Attaining $T=0$ in Figure~\ref{fig:master}B ---non-asympthotically, as noted by \citeauthor{Nernst1924}--- introduces no physical singularity for this process: the rotation of the knob can proceed continuously in the same direction and should conceptually produce the same result: a further decrease in temperature. Mathematically, the equation of state in Figure~\ref{fig:master}B is non-singular at $T=0$, and the analytic continuation of $S(T,x)$ into the negative temperature domain ($T < 0$) is smooth. What physical force could abruptly compel the process path to deviate from its smooth trajectory to $C \to E$? None can be foreseen. For the system under study, the boundary illustrated in Figure~\ref{fig:master}B is merely an illusion arising directly from the standard formulation of the Second Law, revealing a striking inconsistency: the very same Second Law is incapable of demonstrating that $C \to E$ is irreversible or that $C \to D$ is impossible.

\begin{figure*}
  \includegraphics{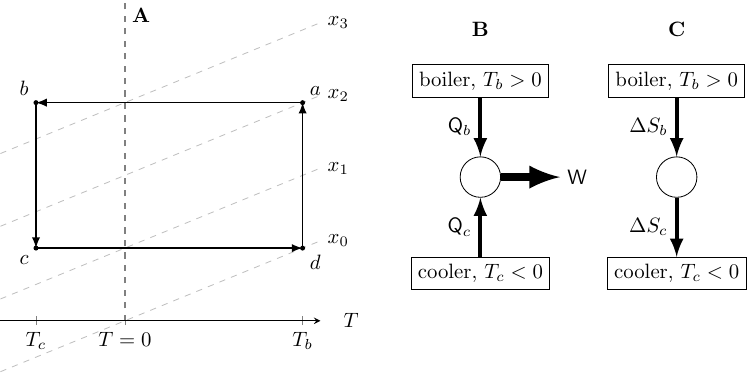}

\caption{The analytic continuation of Figure~\ref{fig:master}B across absolute zero: (A) For an Einsteinian substance, a hypothetical Carnot engine can operate between a positive boiler temperature ($T_b > 0$) and a negative cooler temperature ($T_c < 0$). (B) The corresponding energy flow diagram shows that the engine extracts heat from the boiler and the cooler, causing the negative-temperature reservoir to be thermally ``heated'' as it rejects energy.\citep{Landsberg1977} (C) The corresponding entropy flow diagram shows that entropy is transported from the boiler to the cooler. The dashed $T=0$ axis underscores that a reversible Carnot engine cannot operate at absolute zero itself, maintaining consistency with \citeauthor{Einstein-1913}'s original foundational argument.}  \label{fig:negative}
\end{figure*}

Consistenly, and somewhat unorthodoxly, the Einsteinian substance would admit accessible states at negative temperatures, thereby permitting Carnot engines to operate between a boiler at $T > 0$ and a cooler at $T < 0$ (see Figure~\ref{fig:negative}A). In this scope, a reversible Carnot engine would extract heat from both reservoirs while simultaneously cooling the boiler and heating the cooler (as shown in panel B), while still transporting entropy from the boiler to the cooler (panel C). This does not \emph{literally} violate Planck's formulation of the Second Law, as a third effect---namely, the heat exchanged with the cooler---is enforced.

Yet, this hypothesis directly challenges the fundamental architecture of the Second Law. Formally and historically, the Second Law serves as a structural refinement of the First Law. This ``third effect'' must be strictly energetic, consisting of the heating of a reservoir, which fundamentally implies that the reservoir must receive heat. If absorbing heat from a single reservoir and converting it entirely to work is deemed impossible, absorbing heat from two distinct reservoirs to achieve the same result must be equally restricted. Crucially, while a reservoir at $T < 0$ is formally ``heated'' when it rejects heat, such a conceptual inversion cannot be pre-supposed by the Second Law; doing so would require an \emph{a priori} knowledge of both temperature and entropy, which violates the axiomatic sequence. 

Furthermore, the Einsteinian substance would still remain unable to operate with a reservoir located precisely at $T=0$, as \citeauthor{Einstein-1913}'s original argument remains valid. This restriction is highlighted in Figure~\ref{fig:negative}A by the dashed vertical axis. Consequently, this axis becomes an isolated exclusion region, a feature that deeply challenges any notion of physical continuity across the temperature domain. Alternatively, one might argue that such an engine could be permissible without violating the Second Law on the grounds that a third effect still occurs: the transport of entropy to the reservoir at $T=0$. However, as established above, invoking this defense ---with the third effect being non-energetic--- negates the axiomatic sequence upon which the Second Law is constructed.

It must be noted that the analytic continuation of the isolines in Figure~\ref{fig:master}A below $T=0$ is also perfectly possible, maintaining $(\partial S/\partial T)_x > 0$. Therefore, the state function $S(T,x)$ remains mathematically well-defined for $T < 0$, with the distinct characteristic that the isothermal cross-derivative $(\partial S/\partial x)_T$ would invert its sign as $T$ crosses into the negative domain. Within Nernst's framework, however, the region $T \le 0$ remains fundamentally unattainable by any thermodynamic means due to the intrinsic properties of the substance under study. Specifically, a finite amount of entropy extracted from a boiler at $T_b > 0$ can be transported to some lower positive temperature $0<T_c<T_b$, but it cannot be transported across this accumulation point to a hypothetical cooler located in the range $T_c < 0$. It could only be possible by trespassing the $T\to\infty$ barrier.\citep{Landsberg1977}

It is worth noting that accumulation points of this nature can theoretically occur at any arbitrary temperature. Unlike specific heats or susceptibilities, the Second Law prescribes no inherent constraints on cross-effects, such as the isothermal change of entropy with an external parameter. The profound significance of the Nernst theorem thus lies in its status as a universal property observed across all known homogeneous, finite-density substances. The universal validity of this property at $T=0$ ensures that no auxiliary substance can exist to facilitate the cooling of a system to absolute zero via thermal contact.

\section{Conclusion}
\label{sec:conclusion}

\citet{Blau1996} asked, ``what is the third law trying to tell us?'' If the analysis is directed toward the core of Planck's formulation of the Second Law, Nernst's theorem simply demonstrates that every physical system is fundamentally capable of exchanging entropy with its environment and, consequently, can serve as the low-temperature reservoir of a reversible Carnot engine. Conversely, Einstein's perspective essentially admits the hypothetical existence of a system incapable of playing such role. Within the structural framework of the Second Law, temperature is fundamentally alien to this latter class of systems.

The Second Law originally arose from empirical observations at non-zero temperatures, yet its core statement remains valid irrespective of temperature, as noted by \citet[Chapter~II, \S116]{planck-1897}.\footnote{``The temperature of the reservoirs does not enter into the question. If such a machine [the one that negates the statement] is possible with a reservoir at 1000$^\circ$C, it is also possible with a reservoir at 0$^\circ$C.''} While that remark was not explicitly intended to address the limit $T=0$, the interpretation of Nernst's empirical observations follows this exact logic: they provide the purest evidence that the Second Law holds at any temperature, including the boundary $T=0$. Indeed, it could hardly be otherwise if one acknowledges that temperature is not an independent background coordinate, but a direct consequence of the Second Law and the underlying nature of irreversibility.

In his autobiographical notes, \citet{einstein-1979} famously remarked that ``a theory is the more impressive the greater the simplicity of its premises, the more different kinds of things it relates, and the more extended its area of applicability.'' It is an irony that Einstein wrote these words as a specific tribute to the universal content of thermodynamics. Yet, in his historical rejection of Nernst’s proof, he advocated for a more restricted, less universal Second Law---one that required a ``patch'' in the form of a Third Law to maintain consistency at the boundary. By maintaining the universality of the Second Law irrespective of temperature, we do not merely add Nernst's observations as one of its consequences; we fulfill Einstein’s own desire for a theory of universal applicability. This extension reduces the complexity of the premises by removing the need for a separate, independent postulate to govern the boundary.

The history of science often conflates the order of discovery with the order of logical necessity. One could easily imagine a counter-factual history in which a theorist, working strictly from the axiomatic frameworks of classical thermodynamics, predicted the vanishing of specific heats and the convergence of entropy as a requirement for the consistency of the Second Law. In such a scenario, Nernst’s empirical observations would have been hailed as the brilliant experimental confirmation of an existing theoretical prediction, rather than the discovery of an independent law that only introduces an unnecessary complexity.

In conclusion, the third law is not an independent pillar of thermodynamics, but rather the necessary boundary condition that preserves the logical integrity of the Second Law at its physical limit. The historical and formal debate surrounding this issue has persisted precisely because the third law inherently requires the concepts of temperature and entropy, structures that only find their formal existence and justification through the architecture of the Second Law itself. Acknowledging this structural interdependence does not diminish Nernst’s heat theorem; rather, it elevates the Second Law to that truly \emph{supreme position} \citep{eddington-1915} that Einstein himself so deeply admired.

%% file: main.bbl
%

%% file: history.tex
\section*{Timeline}
\label{sec:timeline}

This work was conceptualized on Jan 12th following a talk at the Departamento de Electrónica y Electromagnetismo from Universidad de Sevilla and discussion therein. The manuscript follows ideas in \doi{10.5281/zenodo.17175342} but turned into a more axiomatic approach and a more explicit description of the havocs associated with Einstein's rebuttal.

The manuscript was originally written in English and refined for grammar an style by \texttt{Gemini}-LLM in collaboration with Universidad de Sevilla. JMM-O thoroughly reviewed, revised and edited the translation.

%% file: main.bbl
\begin{thebibliography}{62}%
\makeatletter
\providecommand \@ifxundefined [1]{%
 \@ifx{#1\undefined}
}%
\providecommand \@ifnum [1]{%
 \ifnum #1\expandafter \@firstoftwo
 \else \expandafter \@secondoftwo
 \fi
}%
\providecommand \@ifx [1]{%
 \ifx #1\expandafter \@firstoftwo
 \else \expandafter \@secondoftwo
 \fi
}%
\providecommand \natexlab [1]{#1}%
\providecommand \enquote  [1]{``#1''}%
\providecommand \bibnamefont  [1]{#1}%
\providecommand \bibfnamefont [1]{#1}%
\providecommand \citenamefont [1]{#1}%
\providecommand \href@noop [0]{\@secondoftwo}%
\providecommand \href [0]{\begingroup \@sanitize@url \@href}%
\providecommand \@href[1]{\@@startlink{#1}\@@href}%
\providecommand \@@href[1]{\endgroup#1\@@endlink}%
\providecommand \@sanitize@url [0]{\catcode `\\12\catcode `\$12\catcode
  `\&12\catcode `\#12\catcode `\^12\catcode `\_12\catcode `\%12\relax}%
\providecommand \@@startlink[1]{}%
\providecommand \@@endlink[0]{}%
\providecommand \url  [0]{\begingroup\@sanitize@url \@url }%
\providecommand \@url [1]{\endgroup\@href {#1}{\urlprefix }}%
\providecommand \urlprefix  [0]{URL }%
\providecommand \Eprint [0]{\href }%
\providecommand \doibase [0]{https://doi.org/}%
\providecommand \selectlanguage [0]{\@gobble}%
\providecommand \bibinfo  [0]{\@secondoftwo}%
\providecommand \bibfield  [0]{\@secondoftwo}%
\providecommand \translation [1]{[#1]}%
\providecommand \BibitemOpen [0]{}%
\providecommand \bibitemStop [0]{}%
\providecommand \bibitemNoStop [0]{.\EOS\space}%
\providecommand \EOS [0]{\spacefactor3000\relax}%
\providecommand \BibitemShut  [1]{\csname bibitem#1\endcsname}%
\let\auto@bib@innerbib\@empty
\bibitem [{\citenamefont {Atkins}(2007)}]{Atkins2007}%
  \BibitemOpen
  \bibfield  {author} {\bibinfo {author} {\bibnamefont {Atkins}, \bibfnamefont
  {P~W}}} (\bibinfo {year} {2007}),\ \href@noop {} {\emph {\bibinfo {title}
  {Four laws that drive the universe}}}\ (\bibinfo  {publisher} {Oxford
  University Press})\BibitemShut {NoStop}%
\bibitem [{\citenamefont {Bazarov}(1971)}]{Bazarov1971}%
  \BibitemOpen
  \bibfield  {author} {\bibinfo {author} {\bibnamefont {Bazarov}, \bibfnamefont
  {Ivan~Pavlovich}}} (\bibinfo {year} {1971}),\ \bibfield  {title} {\enquote
  {\bibinfo {title} {The third law of thermodynamics and its derivations form
  the first and second laws},}\ }\href
  {https://doi.org/10.5281/zenodo.18880680} {\bibfield  {journal} {\bibinfo
  {journal} {Russian Journal of Physical Chemistry}\ }\textbf {\bibinfo
  {volume} {45}},\ \bibinfo {pages} {927--929}}\BibitemShut {NoStop}%
\bibitem [{\citenamefont {Belgiorno}(2003)}]{Belgiorno2003}%
  \BibitemOpen
  \bibfield  {author} {\bibinfo {author} {\bibnamefont {Belgiorno},
  \bibfnamefont {F}}} (\bibinfo {year} {2003}),\ \bibfield  {title} {\enquote
  {\bibinfo {title} {Notes on the third law of thermodynamics: I},}\ }\href
  {https://doi.org/10.1088/0305-4470/36/30/301} {\bibfield  {journal} {\bibinfo
   {journal} {Journal of Physics A: Mathematical and General}\ }\textbf
  {\bibinfo {volume} {36}},\ \bibinfo {pages} {8165}}\BibitemShut {NoStop}%
\bibitem [{\citenamefont {Beretta}\ and\ \citenamefont
  {Gyftopoulos}(2015{\natexlab{a}})}]{Beretta2015c}%
  \BibitemOpen
  \bibfield  {author} {\bibinfo {author} {\bibnamefont {Beretta}, \bibfnamefont
  {Gian~Paolo}}, and\ \bibinfo {author} {\bibfnamefont {Elias~Panayotis}\
  \bibnamefont {Gyftopoulos}}} (\bibinfo {year} {2015}{\natexlab{a}}),\
  \bibfield  {title} {\enquote {\bibinfo {title} {What is heat?}}\ }\href
  {https://doi.org/10.1115/1.4026382/372874} {\bibfield  {journal} {\bibinfo
  {journal} {Journal of Energy Resources Technology, Transactions of the ASME}\
  }\textbf {\bibinfo {volume} {137}},\ 10.1115/1.4026382/372874}\BibitemShut
  {NoStop}%
\bibitem [{\citenamefont {Beretta}\ and\ \citenamefont
  {Gyftopoulos}(2015{\natexlab{b}})}]{Beretta2015e}%
  \BibitemOpen
  \bibfield  {author} {\bibinfo {author} {\bibnamefont {Beretta}, \bibfnamefont
  {Gian~Paolo}}, and\ \bibinfo {author} {\bibfnamefont {Elias~Panayotis}\
  \bibnamefont {Gyftopoulos}}} (\bibinfo {year} {2015}{\natexlab{b}}),\
  \bibfield  {title} {\enquote {\bibinfo {title} {What is the third law?}}\
  }\href {https://doi.org/10.1115/1.4026380/372861} {\bibfield  {journal}
  {\bibinfo  {journal} {Journal of Energy Resources Technology, Transactions of
  the ASME}\ }\textbf {\bibinfo {volume} {137}},\
  10.1115/1.4026380/372861}\BibitemShut {NoStop}%
\bibitem [{\citenamefont {Blau}\ and\ \citenamefont
  {Halfpap}(1996)}]{Blau1996}%
  \BibitemOpen
  \bibfield  {author} {\bibinfo {author} {\bibnamefont {Blau}, \bibfnamefont
  {Steve}}, and\ \bibinfo {author} {\bibfnamefont {Brad}\ \bibnamefont
  {Halfpap}}} (\bibinfo {year} {1996}),\ \bibfield  {title} {\enquote {\bibinfo
  {title} {Question 34. what is the third law of thermodynamics trying to tell
  us?}}\ }\href {https://doi.org/10.1119/1.18284} {\bibfield  {journal}
  {\bibinfo  {journal} {American Journal of Physics}\ }\textbf {\bibinfo
  {volume} {64}},\ \bibinfo {pages} {13--14}}\BibitemShut {NoStop}%
\bibitem [{\citenamefont {Boas}(1960)}]{Boas1960}%
  \BibitemOpen
  \bibfield  {author} {\bibinfo {author} {\bibnamefont {Boas}, \bibfnamefont
  {Mary~Layne}}} (\bibinfo {year} {1960}),\ \bibfield  {title} {\enquote
  {\bibinfo {title} {A point of logic},}\ }\href
  {https://doi.org/10.1119/1.1935931} {\bibfield  {journal} {\bibinfo
  {journal} {American Journal of Physics}\ }\textbf {\bibinfo {volume} {28}},\
  \bibinfo {pages} {675--675}}\BibitemShut {NoStop}%
\bibitem [{\citenamefont {Brown}\ and\ \citenamefont
  {Uffink}(2001)}]{Brown2001}%
  \BibitemOpen
  \bibfield  {author} {\bibinfo {author} {\bibnamefont {Brown}, \bibfnamefont
  {Harvey~R}}, and\ \bibinfo {author} {\bibfnamefont {Jos}\ \bibnamefont
  {Uffink}}} (\bibinfo {year} {2001}),\ \bibfield  {title} {\enquote {\bibinfo
  {title} {The origins of time-asymmetry in thermodynamics: The minus first
  law},}\ }\href {https://doi.org/10.1016/S1355-2198(01)00021-1} {\bibfield
  {journal} {\bibinfo  {journal} {Studies in History and Philosophy of Science
  Part B: Studies in History and Philosophy of Modern Physics}\ }\textbf
  {\bibinfo {volume} {32}},\ \bibinfo {pages} {525--538}}\BibitemShut {NoStop}%
\bibitem [{\citenamefont {Callen}(1985)}]{callen-85}%
  \BibitemOpen
  \bibfield  {author} {\bibinfo {author} {\bibnamefont {Callen}, \bibfnamefont
  {{Herbert Bernard}}}} (\bibinfo {year} {1985}),\ \href@noop {} {\emph
  {\bibinfo {title} {{T}hermodynamics and an {I}ntroduction to
  {T}hermostatistics}}},\ \bibinfo {edition} {2nd}\ ed.\ (\bibinfo  {publisher}
  {John Wiley and Sons})\BibitemShut {NoStop}%
\bibitem [{\citenamefont {Carathéodory}(1909)}]{Caratheodory1909}%
  \BibitemOpen
  \bibfield  {author} {\bibinfo {author} {\bibnamefont {Carathéodory},
  \bibfnamefont {Constantin}}} (\bibinfo {year} {1909}),\ \bibfield  {title}
  {\enquote {\bibinfo {title} {Untersuchungen über die grundlagen der
  thermodynamik},}\ }\href {https://doi.org/10.1007/BF01450409} {\bibfield
  {journal} {\bibinfo  {journal} {Mathematischen Annalen}\ }\textbf {\bibinfo
  {volume} {67}},\ \bibinfo {pages} {355}}\BibitemShut {NoStop}%
\bibitem [{\citenamefont {Carnot}(1824)}]{carnot-1824}%
  \BibitemOpen
  \bibfield  {author} {\bibinfo {author} {\bibnamefont {Carnot}, \bibfnamefont
  {{Nicolas Léonard Sadi}}}} (\bibinfo {year} {1824}),\ \href
  {https://gallica.bnf.fr/ark:/12148/bpt6k29063f} {\emph {\bibinfo {title}
  {R{\'e}flexions sur le puissance motrice du feu, et sur les machines propes
  à d{\`e}veloper cette puissance}}}\ (\bibinfo  {publisher} {Bachelier,
  Paris})\BibitemShut {NoStop}%
\bibitem [{\citenamefont {Chen}\ \emph {et~al.}(2026)\citenamefont {Chen},
  \citenamefont {Pan}, \citenamefont {Zhou},\ and\ \citenamefont
  {Chen}}]{Chen2026}%
  \BibitemOpen
  \bibfield  {author} {\bibinfo {author} {\bibnamefont {Chen}, \bibfnamefont
  {Xiaohang}}, \bibinfo {author} {\bibfnamefont {Yuzhuo}\ \bibnamefont {Pan}},
  \bibinfo {author} {\bibfnamefont {Yinghui}\ \bibnamefont {Zhou}}, and\
  \bibinfo {author} {\bibfnamefont {Jincan}\ \bibnamefont {Chen}}} (\bibinfo
  {year} {2026}),\ \bibfield  {title} {\enquote {\bibinfo {title} {Comment on
  “proof of the nernst theorem”},}\ }\href
  {https://doi.org/10.1140/epjp/s13360-026-07527-6} {\bibfield  {journal}
  {\bibinfo  {journal} {The European Physical Journal Plus}\ }\textbf {\bibinfo
  {volume} {141}},\ \bibinfo {pages} {293--}}\BibitemShut {NoStop}%
\bibitem [{\citenamefont {Clausius}(1854)}]{Clausius1854}%
  \BibitemOpen
  \bibfield  {author} {\bibinfo {author} {\bibnamefont {Clausius},
  \bibfnamefont {Rudolf}}} (\bibinfo {year} {1854}),\ \bibfield  {title}
  {\enquote {\bibinfo {title} {Ueber eine veränderte form des zweiten
  hauptsatzes der mechanischen wärmetheorie},}\ }\href
  {https://doi.org/10.1002/ANDP.18541691202} {\bibfield  {journal} {\bibinfo
  {journal} {Annalen der Physik}\ }\textbf {\bibinfo {volume} {169}},\ \bibinfo
  {pages} {481--506}}\BibitemShut {NoStop}%
\bibitem [{\citenamefont {Debye}(1912)}]{Debye1912}%
  \BibitemOpen
  \bibfield  {author} {\bibinfo {author} {\bibnamefont {Debye}, \bibfnamefont
  {P}}} (\bibinfo {year} {1912}),\ \bibfield  {title} {\enquote {\bibinfo
  {title} {Zur theorie der spezifischen wärmen},}\ }\href
  {https://doi.org/10.1002/ANDP.19123441404} {\bibfield  {journal} {\bibinfo
  {journal} {Annalen der Physik}\ }\textbf {\bibinfo {volume} {344}},\ \bibinfo
  {pages} {789--839}}\BibitemShut {NoStop}%
\bibitem [{\citenamefont {Eddington}(1915)}]{eddington-1915}%
  \BibitemOpen
  \bibfield  {author} {\bibinfo {author} {\bibnamefont {Eddington},
  \bibfnamefont {Arthur~Stanley}}} (\bibinfo {year} {1915}),\ \href@noop {}
  {\emph {\bibinfo {title} {The Nature of the Physical World}}}\ (\bibinfo
  {publisher} {Kessinger Publishing, LLC (2010)})\BibitemShut {NoStop}%
\bibitem [{\citenamefont {Ehrlich}(1981)}]{Ehrlich1981}%
  \BibitemOpen
  \bibfield  {author} {\bibinfo {author} {\bibnamefont {Ehrlich}, \bibfnamefont
  {Philip}}} (\bibinfo {year} {1981}),\ \bibfield  {title} {\enquote {\bibinfo
  {title} {The concept of temperature and its dependence on the laws of
  thermodynamics},}\ }\href {https://doi.org/10.1119/1.12448} {\bibfield
  {journal} {\bibinfo  {journal} {American Journal of Physics}\ }\textbf
  {\bibinfo {volume} {49}},\ \bibinfo {pages} {622--632}}\BibitemShut {NoStop}%
\bibitem [{\citenamefont {Einstein}(1907)}]{Einstein1907}%
  \BibitemOpen
  \bibfield  {author} {\bibinfo {author} {\bibnamefont {Einstein},
  \bibfnamefont {A}}} (\bibinfo {year} {1907}),\ \bibfield  {title} {\enquote
  {\bibinfo {title} {Die plancksche theorie der strahlung und die theorie der
  spezifischen wärme},}\ }\href {https://doi.org/10.1002/ANDP.19063270110}
  {\bibfield  {journal} {\bibinfo  {journal} {Annalen der Physik}\ }\textbf
  {\bibinfo {volume} {327}},\ \bibinfo {pages} {180--190}}\BibitemShut
  {NoStop}%
\bibitem [{\citenamefont {Einstein}(1913)}]{Einstein-1913}%
  \BibitemOpen
  \bibfield  {author} {\bibinfo {author} {\bibnamefont {Einstein},
  \bibfnamefont {Albert}}} (\bibinfo {year} {1913}),\ in\ \href
  {https://dipot.ulb.ac.be/dspace/bitstream/2013/234808/3/DL2053602_000_f.pdf}
  {\emph {\bibinfo {booktitle} {La structure de la matière. Rapports et
  discussions du Conseil de physique tenu à Bruxelles du 27 au 31 octobre
  1913}}},\ \bibinfo {editor} {edited by\ \bibinfo {editor} {\bibfnamefont
  {Institut~International}\ \bibnamefont {de~Physique~Solvay}}},\ \bibinfo
  {organization} {Commission administrative de l'Institut Solvay}\ (\bibinfo
  {publisher} {Gauthier-Villars et Cie})\ pp.\ \bibinfo {pages}
  {293--298}\BibitemShut {NoStop}%
\bibitem [{\citenamefont {Einstein}(1979)}]{einstein-1979}%
  \BibitemOpen
  \bibfield  {author} {\bibinfo {author} {\bibnamefont {Einstein},
  \bibfnamefont {Albert}}} (\bibinfo {year} {1979}),\ \href@noop {} {\emph
  {\bibinfo {title} {Autobiographical Notes. A Centennial Edition}}},\ edited
  by\ \bibinfo {editor} {\bibfnamefont {Paul~Arthur}\ \bibnamefont {Schilpp}}\
  (\bibinfo  {publisher} {Open Court Publishing Company})\BibitemShut {NoStop}%
\bibitem [{\citenamefont {Epstein}(1937)}]{epstein-1937}%
  \BibitemOpen
  \bibfield  {author} {\bibinfo {author} {\bibnamefont {Epstein}, \bibfnamefont
  {{Paul Sophus}}}} (\bibinfo {year} {1937}),\ \href@noop {} {\emph {\bibinfo
  {title} {A Textbook of Thermodynamics}}}\ (\bibinfo  {publisher} {John Wiley
  and Sons})\BibitemShut {NoStop}%
\bibitem [{\citenamefont {Falk}(1959)}]{Falk1959}%
  \BibitemOpen
  \bibfield  {author} {\bibinfo {author} {\bibnamefont {Falk}, \bibfnamefont
  {Gottfried}}} (\bibinfo {year} {1959}),\ \bibfield  {title} {\enquote
  {\bibinfo {title} {Third law of thermodynamics},}\ }\href
  {https://doi.org/10.1103/PhysRev.115.249} {\bibfield  {journal} {\bibinfo
  {journal} {Physical Review}\ }\textbf {\bibinfo {volume} {115}},\ \bibinfo
  {pages} {249--253}}\BibitemShut {NoStop}%
\bibitem [{\citenamefont {Gyftopoulos}\ and\ \citenamefont
  {Beretta}(2005)}]{gyftopoulos-05}%
  \BibitemOpen
  \bibfield  {author} {\bibinfo {author} {\bibnamefont {Gyftopoulos},
  \bibfnamefont {Elias~Panayotis}}, and\ \bibinfo {author} {\bibfnamefont
  {Gian~Paolo}\ \bibnamefont {Beretta}}} (\bibinfo {year} {2005}),\ \href@noop
  {} {\emph {\bibinfo {title} {Thermodynamics: foundations and applications}}}\
  (\bibinfo  {publisher} {Dover})\BibitemShut {NoStop}%
\bibitem [{\citenamefont {Hatsopoulos}\ and\ \citenamefont
  {Keenan}(1965)}]{Hatsopoulos1965}%
  \BibitemOpen
  \bibfield  {author} {\bibinfo {author} {\bibnamefont {Hatsopoulos},
  \bibfnamefont {George~N}}, and\ \bibinfo {author} {\bibfnamefont {Joseph~H.}\
  \bibnamefont {Keenan}}} (\bibinfo {year} {1965}),\ \href@noop {} {\emph
  {\bibinfo {title} {Principles of General Thermodynamics}}}\ (\bibinfo
  {publisher} {John Wiley and Sons})\BibitemShut {NoStop}%
\bibitem [{\citenamefont {Heidrich}(2016)}]{Heidrich2016}%
  \BibitemOpen
  \bibfield  {author} {\bibinfo {author} {\bibnamefont {Heidrich},
  \bibfnamefont {Matthias}}} (\bibinfo {year} {2016}),\ \bibfield  {title}
  {\enquote {\bibinfo {title} {Bounded energy exchange as an alternative to the
  third law of thermodynamics},}\ }\href
  {https://doi.org/10.1016/J.AOP.2016.07.031} {\bibfield  {journal} {\bibinfo
  {journal} {Annals of Physics}\ }\textbf {\bibinfo {volume} {373}},\ \bibinfo
  {pages} {665--681}}\BibitemShut {NoStop}%
\bibitem [{\citenamefont {Kestin}(1968)}]{kestin-1968ii}%
  \BibitemOpen
  \bibfield  {author} {\bibinfo {author} {\bibnamefont {Kestin}, \bibfnamefont
  {Joseph}}} (\bibinfo {year} {1968}),\ \href@noop {} {\emph {\bibinfo {title}
  {A course in Thermodynamics}}},\ Vol.~\bibinfo {volume} {II}\ (\bibinfo
  {publisher} {Blaisdell Pub. Co.})\BibitemShut {NoStop}%
\bibitem [{\citenamefont {Klimenko}(2012)}]{Klimenko2012}%
  \BibitemOpen
  \bibfield  {author} {\bibinfo {author} {\bibnamefont {Klimenko},
  \bibfnamefont {A~Y}}} (\bibinfo {year} {2012}),\ \bibfield  {title} {\enquote
  {\bibinfo {title} {Teaching the third law of thermodynamics},}\ }\href
  {https://doi.org/10.2174/1874396X01206010001} {\bibfield  {journal} {\bibinfo
   {journal} {The Open Thermodynamics Journal}\ }\textbf {\bibinfo {volume}
  {6}},\ \bibinfo {pages} {1--14}}\BibitemShut {NoStop}%
\bibitem [{\citenamefont {Kox}(2006)}]{Kox2006}%
  \BibitemOpen
  \bibfield  {author} {\bibinfo {author} {\bibnamefont {Kox}, \bibfnamefont
  {A~J}}} (\bibinfo {year} {2006}),\ \bibfield  {title} {\enquote {\bibinfo
  {title} {Confusion and clarification: Albert einstein and walther nernst's
  heat theorem, 1911–1916},}\ }\href
  {https://doi.org/10.1016/J.SHPSB.2005.10.001} {\bibfield  {journal} {\bibinfo
   {journal} {Studies in History and Philosophy of Science Part B: Studies in
  History and Philosophy of Modern Physics}\ }\textbf {\bibinfo {volume}
  {37}},\ \bibinfo {pages} {101--114}}\BibitemShut {NoStop}%
\bibitem [{\citenamefont {Landau}\ and\ \citenamefont
  {Lifshitz}(1968)}]{landau-lifshitz-1968}%
  \BibitemOpen
  \bibfield  {author} {\bibinfo {author} {\bibnamefont {Landau}, \bibfnamefont
  {{Lev Davidovich}}}, and\ \bibinfo {author} {\bibfnamefont {{E. M.}}\
  \bibnamefont {Lifshitz}}} (\bibinfo {year} {1968}),\ \href@noop {} {\emph
  {\bibinfo {title} {Statistical Physics}}}\ (\bibinfo  {publisher} {Pergamon
  Press})\BibitemShut {NoStop}%
\bibitem [{\citenamefont {Landsberg}(1956)}]{Landsberg1956}%
  \BibitemOpen
  \bibfield  {author} {\bibinfo {author} {\bibnamefont {Landsberg},
  \bibfnamefont {Peter~Theodore}}} (\bibinfo {year} {1956}),\ \bibfield
  {title} {\enquote {\bibinfo {title} {Foundations of thermodynamics},}\ }\href
  {https://doi.org/10.1103/RevModPhys.28.363} {\bibfield  {journal} {\bibinfo
  {journal} {Reviews of Modern Physics}\ }\textbf {\bibinfo {volume} {28}},\
  \bibinfo {pages} {363}}\BibitemShut {NoStop}%
\bibitem [{\citenamefont {Landsberg}(1977)}]{Landsberg1977}%
  \BibitemOpen
  \bibfield  {author} {\bibinfo {author} {\bibnamefont {Landsberg},
  \bibfnamefont {Peter~Theodore}}} (\bibinfo {year} {1977}),\ \bibfield
  {title} {\enquote {\bibinfo {title} {Heat engines and heat pumps at positive
  and negative absolute temperatures},}\ }\href
  {https://doi.org/10.1088/0305-4470/10/10/011} {\bibfield  {journal} {\bibinfo
   {journal} {Journal of Physics A: Mathematical and General}\ }\textbf
  {\bibinfo {volume} {10}},\ \bibinfo {pages} {1773--1770}}\BibitemShut
  {NoStop}%
\bibitem [{\citenamefont {Landsberg}(1989)}]{Landsberg1989}%
  \BibitemOpen
  \bibfield  {author} {\bibinfo {author} {\bibnamefont {Landsberg},
  \bibfnamefont {Peter~Theodore}}} (\bibinfo {year} {1989}),\ \bibfield
  {title} {\enquote {\bibinfo {title} {A comment on nernst's theorem},}\ }\href
  {https://doi.org/10.1088/0305-4470/22/1/021} {\bibfield  {journal} {\bibinfo
  {journal} {Journal of Physics A: Mathematical and General}\ }\textbf
  {\bibinfo {volume} {22}},\ \bibinfo {pages} {139}}\BibitemShut {NoStop}%
\bibitem [{\citenamefont {Landsberg}(1997)}]{landsberg-ajp-97}%
  \BibitemOpen
  \bibfield  {author} {\bibinfo {author} {\bibnamefont {Landsberg},
  \bibfnamefont {Peter~Theodore}}} (\bibinfo {year} {1997}),\ \bibfield
  {title} {\enquote {\bibinfo {title} {Answer to question 34. what is the third
  law of thermodynamics trying to tell us?}}\ }\href
  {https://doi.org/10.1119/1.18284} {\bibfield  {journal} {\bibinfo  {journal}
  {American Journal of Physics}\ }\textbf {\bibinfo {volume} {65}},\ \bibinfo
  {pages} {269--270}}\BibitemShut {NoStop}%
\bibitem [{\citenamefont {Lavis}\ and\ \citenamefont
  {Frigg}(2025)}]{Lavis2025}%
  \BibitemOpen
  \bibfield  {author} {\bibinfo {author} {\bibnamefont {Lavis}, \bibfnamefont
  {D~A}}, and\ \bibinfo {author} {\bibfnamefont {Roman.}\ \bibnamefont
  {Frigg}}} (\bibinfo {year} {2025}),\ \href
  {https://doi.org/10.1007/978-3-031-77948-0} {\emph {\bibinfo {title} {The
  fundamentals of thermodynamics}}}\ (\bibinfo  {publisher}
  {Springer})\BibitemShut {NoStop}%
\bibitem [{\citenamefont {Liboff}(1994)}]{liboff-physicsessays-94}%
  \BibitemOpen
  \bibfield  {author} {\bibinfo {author} {\bibnamefont {Liboff}, \bibfnamefont
  {R~L}}} (\bibinfo {year} {1994}),\ \bibfield  {title} {\enquote {\bibinfo
  {title} {The carnot engine in the vicinity of 0k},}\ }\href
  {https://doi.org/10.4006/1.3029121} {\bibfield  {journal} {\bibinfo
  {journal} {Physics Essays}\ }\textbf {\bibinfo {volume} {7}},\ \bibinfo
  {pages} {95--98}}\BibitemShut {NoStop}%
\bibitem [{\citenamefont {Lieb}\ and\ \citenamefont
  {Yngvason}(1999)}]{Lieb1999}%
  \BibitemOpen
  \bibfield  {author} {\bibinfo {author} {\bibnamefont {Lieb}, \bibfnamefont
  {Elliott~H}}, and\ \bibinfo {author} {\bibfnamefont {Jakob}\ \bibnamefont
  {Yngvason}}} (\bibinfo {year} {1999}),\ \bibfield  {title} {\enquote
  {\bibinfo {title} {The physics and mathematics of the second law of
  thermodynamics},}\ }\href {https://doi.org/10.1016/S0370-1573(98)00082-9}
  {\bibfield  {journal} {\bibinfo  {journal} {Physics Report}\ }\textbf
  {\bibinfo {volume} {310}},\ \bibinfo {pages} {1--96}}\BibitemShut {NoStop}%
\bibitem [{\citenamefont {Mafé}\ and\ \citenamefont {de~la
  Rubia}(1998)}]{mafe-ajp-98}%
  \BibitemOpen
  \bibfield  {author} {\bibinfo {author} {\bibnamefont {Mafé}, \bibfnamefont
  {Salvador}}, and\ \bibinfo {author} {\bibfnamefont {Juan}\ \bibnamefont
  {de~la Rubia}}} (\bibinfo {year} {1998}),\ \bibfield  {title} {\enquote
  {\bibinfo {title} {Answer to question 34. what is the third law of
  thermodynamics trying to tell us?}}\ }\href {https://doi.org/10.1119/1.18869}
  {\bibfield  {journal} {\bibinfo  {journal} {American Journal of Physics}\
  }\textbf {\bibinfo {volume} {66}},\ \bibinfo {pages} {277}}\BibitemShut
  {NoStop}%
\bibitem [{\citenamefont {Martín-Olalla}\ and\ \citenamefont
  {Luna}(2003)}]{Martin-Olalla2003b}%
  \BibitemOpen
  \bibfield  {author} {\bibinfo {author} {\bibnamefont {Martín-Olalla},
  \bibfnamefont {JM}}, and\ \bibinfo {author} {\bibfnamefont {A.~Rey~De}\
  \bibnamefont {Luna}}} (\bibinfo {year} {2003}),\ \bibfield  {title} {\enquote
  {\bibinfo {title} {Universal restrictions to the conversion of heat into work
  derived from the analysis of the nernst theorem as a uniform limit},}\ }\href
  {https://doi.org/10.1088/0305-4470/36/29/303} {\bibfield  {journal} {\bibinfo
   {journal} {Journal of Physics A: Mathematical and General}\ }\textbf
  {\bibinfo {volume} {36}},\ \bibinfo {pages} {7909--7921}}\BibitemShut
  {NoStop}%
\bibitem [{\citenamefont {Martín-Olalla}(2024)}]{Martin-Olalla2024f}%
  \BibitemOpen
  \bibfield  {author} {\bibinfo {author} {\bibnamefont {Martín-Olalla},
  \bibfnamefont {José~María}}} (\bibinfo {year} {2024}),\ \bibfield  {title}
  {\enquote {\bibinfo {title} {A concise rebuttal on a concise proof of the
  equivalence of the nernst theorem and the heat capacity statement of the
  third law of thermodynamics},}\ }\href
  {https://doi.org/10.1142/S0217732324750038} {\bibfield  {journal} {\bibinfo
  {journal} {Modern Physics Letters A}\ }\textbf {\bibinfo {volume} {39}},\
  \bibinfo {pages} {2475003}}\BibitemShut {NoStop}%
\bibitem [{\citenamefont
  {Martín-Olalla}(2025{\natexlab{a}})}]{Martin-Olalla2025c}%
  \BibitemOpen
  \bibfield  {author} {\bibinfo {author} {\bibnamefont {Martín-Olalla},
  \bibfnamefont {José-María}}} (\bibinfo {year} {2025}{\natexlab{a}}),\
  \bibfield  {title} {\enquote {\bibinfo {title} {Proof of the nernst
  theorem},}\ }\href {https://doi.org/10.1140/epjp/s13360-025-06503-w}
  {\bibfield  {journal} {\bibinfo  {journal} {The European Physical Journal
  Plus}\ }\textbf {\bibinfo {volume} {140}},\ \bibinfo {pages}
  {528}}\BibitemShut {NoStop}%
\bibitem [{\citenamefont
  {Martín-Olalla}(2025{\natexlab{b}})}]{Martin-Olalla2025d}%
  \BibitemOpen
  \bibfield  {author} {\bibinfo {author} {\bibnamefont {Martín-Olalla},
  \bibfnamefont {José-María}}} (\bibinfo {year} {2025}{\natexlab{b}}),\
  \bibfield  {title} {\enquote {\bibinfo {title} {Thermal stability originates
  the vanishing of the specific heats at absolute zero},}\ }\href
  {https://doi.org/10.1088/1402-4896/ae22a5} {\bibfield  {journal} {\bibinfo
  {journal} {Physica Scripta}\ }\textbf {\bibinfo {volume} {100}},\ \bibinfo
  {pages} {125206}}\BibitemShut {NoStop}%
\bibitem [{\citenamefont {Martín-Olalla}(2026)}]{Martin-Olalla2026b}%
  \BibitemOpen
  \bibfield  {author} {\bibinfo {author} {\bibnamefont {Martín-Olalla},
  \bibfnamefont {José~María}}} (\bibinfo {year} {2026}),\ \bibfield  {title}
  {\enquote {\bibinfo {title} {Characterizing the carnot cycle at absolute
  zero: a reply to "comment on 'proof of the nernst theorem'"},}\ }\href
  {https://doi.org/10.5281/zenodo.19127423} {\bibfield  {journal} {\bibinfo
  {journal} {zenodo}\ }10.5281/zenodo.19127423}\BibitemShut {NoStop}%
\bibitem [{\citenamefont {Massanes}\ and\ \citenamefont
  {Oppenheim}(2017)}]{Masanes2017}%
  \BibitemOpen
  \bibfield  {author} {\bibinfo {author} {\bibnamefont {Massanes},
  \bibfnamefont {Lluís}}, and\ \bibinfo {author} {\bibfnamefont {Jonathan}\
  \bibnamefont {Oppenheim}}} (\bibinfo {year} {2017}),\ \bibfield  {title}
  {\enquote {\bibinfo {title} {A general derivation and quantification of the
  third law of thermodynamics},}\ }\href {https://doi.org/10.1038/ncomms14538}
  {\bibfield  {journal} {\bibinfo  {journal} {Nature Communications}\ }\textbf
  {\bibinfo {volume} {8}},\ \bibinfo {pages} {1--7}}\BibitemShut {NoStop}%
\bibitem [{\citenamefont {Nernst}(1906)}]{Nernst1906}%
  \BibitemOpen
  \bibfield  {author} {\bibinfo {author} {\bibnamefont {Nernst}, \bibfnamefont
  {W}}} (\bibinfo {year} {1906}),\ \bibfield  {title} {\enquote {\bibinfo
  {title} {Ueber die berechnung chemischer gleichgewichte aus thermischen
  messungen},}\ }\href {https://eudml.org/doc/58630} {\bibfield  {journal}
  {\bibinfo  {journal} {Nachrichten von der Gesellschaft der Wissenschaften zu
  Göttingen, Mathematisch-Physikalische Klasse}\ }\textbf {\bibinfo {volume}
  {1906}},\ \bibinfo {pages} {1--40}}\BibitemShut {NoStop}%
\bibitem [{\citenamefont {Nernst}(1912{\natexlab{a}})}]{Nernst1912}%
  \BibitemOpen
  \bibfield  {author} {\bibinfo {author} {\bibnamefont {Nernst}, \bibfnamefont
  {Walther~Hermann}}} (\bibinfo {year} {1912}{\natexlab{a}}),\ \bibfield
  {title} {\enquote {\bibinfo {title} {Thermodynamik und spezifische wärme},}\
  }\href {https://edoc.bbaw.de/files/5004/BBAW_SB_1912_TB1_S134_140.pdf}
  {\bibfield  {journal} {\bibinfo  {journal} {Sitzber. preuss. Akad. Wiss.
  Physik-math}\ }\textbf {\bibinfo {volume} {K}},\ \bibinfo {pages}
  {134}}\BibitemShut {NoStop}%
\bibitem [{\citenamefont {Nernst}(1912{\natexlab{b}})}]{Nernst-1912}%
  \BibitemOpen
  \bibfield  {author} {\bibinfo {author} {\bibnamefont {Nernst}, \bibfnamefont
  {Walther~Hermann}}} (\bibinfo {year} {1912}{\natexlab{b}}),\ \bibfield
  {title} {\enquote {\bibinfo {title} {Thermodynamik und spezifische
  {W}ärme},}\ }\href {https://www.archive.org/details/mobot31753002089602}
  {\bibfield  {journal} {\bibinfo  {journal} {Sitzber. preuss. {A}kad. {W}iss.
  Physik-math}\ }\textbf {\bibinfo {volume} {K}},\ \bibinfo {pages}
  {134}}\BibitemShut {NoStop}%
\bibitem [{\citenamefont {Nernst}(1924)}]{Nernst1924}%
  \BibitemOpen
  \bibfield  {author} {\bibinfo {author} {\bibnamefont {Nernst}, \bibfnamefont
  {Walther~Hermann}}} (\bibinfo {year} {1924}),\ \href@noop {} {\emph {\bibinfo
  {title} {The New Heat Theorem: Its Foundations in Theory and
  Experimental}}},\ \bibinfo {edition} {2nd}\ ed.\ (\bibinfo  {publisher}
  {Dover})\BibitemShut {NoStop}%
\bibitem [{\citenamefont {Oppenheim}(1989)}]{Oppenheim1989}%
  \BibitemOpen
  \bibfield  {author} {\bibinfo {author} {\bibnamefont {Oppenheim},
  \bibfnamefont {I}}} (\bibinfo {year} {1989}),\ \bibfield  {title} {\enquote
  {\bibinfo {title} {Comment on 'an equivalent theorem of the nernst
  theorem'},}\ }\href {https://doi.org/10.1088/0305-4470/22/1/022} {\bibfield
  {journal} {\bibinfo  {journal} {Journal of Physics A: Mathematical and
  General}\ }\textbf {\bibinfo {volume} {22}},\ \bibinfo {pages}
  {143}}\BibitemShut {NoStop}%
\bibitem [{\citenamefont {Planck}(1897)}]{planck-1897}%
  \BibitemOpen
  \bibfield  {author} {\bibinfo {author} {\bibnamefont {Planck}, \bibfnamefont
  {Max}}} (\bibinfo {year} {1897}),\ \href
  {http://www.gutenberg.org/ebooks/31564} {\emph {\bibinfo {title} {Vorlesungen
  {\"u}ber {T}hermodynamik}}}\ (\bibinfo  {publisher} {Veit und
  Comp.})\BibitemShut {NoStop}%
\bibitem [{\citenamefont {Planck}(1921)}]{Planck1921}%
  \BibitemOpen
  \bibfield  {author} {\bibinfo {author} {\bibnamefont {Planck}, \bibfnamefont
  {Max}}} (\bibinfo {year} {1921}),\ \href@noop {} {\emph {\bibinfo {title}
  {Vorlesungen ueber Thermodynamik}}},\ \bibinfo {edition} {6th}\ ed.\
  (\bibinfo  {publisher} {W. de Gruyter})\BibitemShut {NoStop}%
\bibitem [{\citenamefont {Poincaré}(1892)}]{Poincare1892}%
  \BibitemOpen
  \bibfield  {author} {\bibinfo {author} {\bibnamefont {Poincaré},
  \bibfnamefont {Henri}}} (\bibinfo {year} {1892}),\ \href@noop {} {\emph
  {\bibinfo {title} {Thermodynamique}}}\ (\bibinfo  {publisher}
  {Gauthier-Villars Rep Forgotten Books (2018)})\BibitemShut {NoStop}%
\bibitem [{\citenamefont {Rose-Innes}(1999)}]{roseinnes-ajp-99}%
  \BibitemOpen
  \bibfield  {author} {\bibinfo {author} {\bibnamefont {Rose-Innes},
  \bibfnamefont {C}}} (\bibinfo {year} {1999}),\ \bibfield  {title} {\enquote
  {\bibinfo {title} {Answer to question 34. what is the third law of
  thermodynamics trying to tell us?}}\ }\href {https://doi.org/10.1119/1.19240}
  {\bibfield  {journal} {\bibinfo  {journal} {American Journal of Physics}\
  }\textbf {\bibinfo {volume} {67}},\ \bibinfo {pages} {27}}\BibitemShut
  {NoStop}%
\bibitem [{\citenamefont {Simon}(1951)}]{Simon1951}%
  \BibitemOpen
  \bibfield  {author} {\bibinfo {author} {\bibnamefont {Simon}, \bibfnamefont
  {F~E}}} (\bibinfo {year} {1951}),\ \bibfield  {title} {\enquote {\bibinfo
  {title} {Some considerations concerning nernst’s theorem},}\ }\href
  {https://doi.org/10.1515/ZNA-1951-0717} {\bibfield  {journal} {\bibinfo
  {journal} {Zeitschrift fur Naturforschung - Section A Journal of Physical
  Sciences}\ }\textbf {\bibinfo {volume} {6}},\ \bibinfo {pages}
  {397--400b}}\BibitemShut {NoStop}%
\bibitem [{\citenamefont {Sommerfeld}(1928)}]{Sommerfeld1928}%
  \BibitemOpen
  \bibfield  {author} {\bibinfo {author} {\bibnamefont {Sommerfeld},
  \bibfnamefont {A}}} (\bibinfo {year} {1928}),\ \bibfield  {title} {\enquote
  {\bibinfo {title} {Zur elektronentheorie der metalle auf grund der fermischen
  statistik - i. teil: Allgemeines, strömungs- und austrittsvorgänge},}\
  }\href {https://doi.org/10.1007/BF01391052} {\bibfield  {journal} {\bibinfo
  {journal} {Zeitschrift für Physik}\ }\textbf {\bibinfo {volume} {47}},\
  \bibinfo {pages} {1--32}}\BibitemShut {NoStop}%
\bibitem [{\citenamefont {Su}\ and\ \citenamefont {Chen}(2022)}]{Su2022}%
  \BibitemOpen
  \bibfield  {author} {\bibinfo {author} {\bibnamefont {Su}, \bibfnamefont
  {Shanhe}}, and\ \bibinfo {author} {\bibfnamefont {Jincan}\ \bibnamefont
  {Chen}}} (\bibinfo {year} {2022}),\ \bibfield  {title} {\enquote {\bibinfo
  {title} {A concise proof of the equivalence of the nernst theorem and the
  heat capacity statement of the third law of thermodynamics},}\ }\href
  {https://doi.org/10.1142/S0217732322502467} {\bibfield  {journal} {\bibinfo
  {journal} {Modern Physics Letters A}\ }\textbf {\bibinfo {volume} {37}},\
  \bibinfo {pages} {2250246}}\BibitemShut {NoStop}%
\bibitem [{\citenamefont {Tasaki}\ and\ \citenamefont
  {Paquette}(2026)}]{Tasaki2026}%
  \BibitemOpen
  \bibfield  {author} {\bibinfo {author} {\bibnamefont {Tasaki}, \bibfnamefont
  {Hal}}, and\ \bibinfo {author} {\bibfnamefont {Glenn}\ \bibnamefont
  {Paquette}}} (\bibinfo {year} {2026}),\ \href@noop {} {\emph {\bibinfo
  {title} {Thermodynamics: a modern approach}}}\ (\bibinfo  {publisher} {Oxford
  University Press})\BibitemShut {NoStop}%
\bibitem [{\citenamefont {Thomson}(1848)}]{thomson-1848}%
  \BibitemOpen
  \bibfield  {author} {\bibinfo {author} {\bibnamefont {Thomson}, \bibfnamefont
  {William}}} (\bibinfo {year} {1848}),\ \bibfield  {title} {\enquote {\bibinfo
  {title} {On an absolute thermometric scale founded on carnot's theory of the
  motive power of heat, and calculted from regnault's observations},}\ }\href
  {http://zapatopi.net/kelvin/papers/on_an_absolute_thermometric_scale.html}
  {\bibfield  {journal} {\bibinfo  {journal} {Proc. Cam. Phil. Soc.}\ }\textbf
  {\bibinfo {volume} {1}},\ \bibinfo {pages} {69}}\BibitemShut {NoStop}%
\bibitem [{\citenamefont {Thomson}(1853)}]{thomson-1853}%
  \BibitemOpen
  \bibfield  {author} {\bibinfo {author} {\bibnamefont {Thomson}, \bibfnamefont
  {William}}} (\bibinfo {year} {1853}),\ \bibfield  {title} {\enquote {\bibinfo
  {title} {On the dynamical theory of heat, with numerical results deduced from
  mr joule's equivalent of a thermal unit and m. regnault's observations on
  steam},}\ }\href {https://www.biodiversitylibrary.org/item/126047} {\bibfield
   {journal} {\bibinfo  {journal} {Trans. Roy. Soc. Edinb.}\ }\textbf {\bibinfo
  {volume} {20}},\ \bibinfo {pages} {261}}\BibitemShut {NoStop}%
\bibitem [{\citenamefont {Uffink}(2007)}]{Uffink2007}%
  \BibitemOpen
  \bibfield  {author} {\bibinfo {author} {\bibnamefont {Uffink}, \bibfnamefont
  {Jos}}} (\bibinfo {year} {2007}),\ \enquote {\bibinfo {title} {Compendium of
  the foundations of classical statistical physics},}\ in\ \href
  {https://doi.org/10.1016/B978-044451560-5/50012-9} {\emph {\bibinfo
  {booktitle} {Philosophy of Physics}}}\ (\bibinfo  {publisher} {Elsevier})\
  pp.\ \bibinfo {pages} {923--1074}\BibitemShut {NoStop}%
\bibitem [{\citenamefont {Uffink}(2017)}]{Uffink2017}%
  \BibitemOpen
  \bibfield  {author} {\bibinfo {author} {\bibnamefont {Uffink}, \bibfnamefont
  {Jos}}} (\bibinfo {year} {2017}),\ \bibfield  {title} {\enquote {\bibinfo
  {title} {Massanes and oppenheim on the third law of thermodynamics},}\ }\href
  {https://doi.org/10.1007/s10701-017-0095-2} {\bibfield  {journal} {\bibinfo
  {journal} {Foundations of Physics}\ }\textbf {\bibinfo {volume} {47}},\
  \bibinfo {pages} {871--872}}\BibitemShut {NoStop}%
\bibitem [{\citenamefont {Wheeler}(1991)}]{Wheeler1991}%
  \BibitemOpen
  \bibfield  {author} {\bibinfo {author} {\bibnamefont {Wheeler}, \bibfnamefont
  {John~C}}} (\bibinfo {year} {1991}),\ \bibfield  {title} {\enquote {\bibinfo
  {title} {Nonequivalence of the nernst-simon and unattainability statements of
  the third law of thermodynamics},}\ }\href
  {https://doi.org/10.1103/PhysRevA.43.5289} {\bibfield  {journal} {\bibinfo
  {journal} {Physical Review A}\ }\textbf {\bibinfo {volume} {43}},\ \bibinfo
  {pages} {5289}}\BibitemShut {NoStop}%
\bibitem [{\citenamefont {Yan}\ and\ \citenamefont {Chen}(1988)}]{Yan1988}%
  \BibitemOpen
  \bibfield  {author} {\bibinfo {author} {\bibnamefont {Yan}, \bibfnamefont
  {Zijun}}, and\ \bibinfo {author} {\bibfnamefont {Jincan}\ \bibnamefont
  {Chen}}} (\bibinfo {year} {1988}),\ \bibfield  {title} {\enquote {\bibinfo
  {title} {An equivalent theorem of the nernst theorem},}\ }\href
  {https://doi.org/10.1088/0305-4470/21/13/006} {\bibfield  {journal} {\bibinfo
   {journal} {Journal of Physics A: Mathematical and General}\ }\textbf
  {\bibinfo {volume} {21}},\ \bibinfo {pages} {707--709}}\BibitemShut {NoStop}%
\bibitem [{\citenamefont {Young}(1975)}]{Young1975}%
  \BibitemOpen
  \bibfield  {author} {\bibinfo {author} {\bibnamefont {Young}, \bibfnamefont
  {David~A}}} (\bibinfo {year} {1975}),\ \href
  {https://www.osti.gov/servlets/purl/4010212} {\emph {\bibinfo {title} {Phase
  Diagrams of the Elements}}},\ \bibinfo {type} {Tech. Rep.}\ (\bibinfo
  {institution} {Lawrence Livermore Laboratory UCLA})\BibitemShut {NoStop}%
\end{thebibliography}
